\documentclass[11pt,english,aps,superscriptaddress]{revtex4}
\usepackage[latin9]{inputenc}
\usepackage{babel}
\usepackage{amsmath}
\usepackage{amssymb}
\usepackage{graphicx}
\usepackage[colorlinks,linkcolor=red,anchorcolor=blue,citecolor=blue]{hyperref}
\makeatletter

\usepackage{amsthm}
\usepackage{amsfonts}
\usepackage{epsfig}
\usepackage{babel}
\usepackage{color}
\usepackage{framed}
\usepackage{float}
\usepackage{array}

\makeatother
\begin{document}
  \title{Dynamical spontaneous scalarization in Einstein-Maxwell-scalar models in anti-de Sitter spacetime}
  \author{Wen-Kun Luo}
  \email{luowk@stu2020.jnu.edu.cn}
  
  \address{\textit{Department of Physics and Siyuan Laboratory, Jinan University, Guangzhou 510632, China }} 
  \author{Cheng-Yong Zhang}
  \email{zhangcy@email.jnu.edu.cn}
  
  \address{\textit{Department of Physics and Siyuan Laboratory, Jinan University, Guangzhou 510632, China }}
  \author{Peng Liu}
  \email{phylp@email.jnu.edu.cn}
  
  \address{\textit{Department of Physics and Siyuan Laboratory, Jinan University, Guangzhou 510632, China }}
  \author{Chao Niu}
  \email{niuchaophy@gmail.com}
  
  \address{\textit{Department of Physics and Siyuan Laboratory, Jinan University, Guangzhou 510632, China }}
  \author{Bin Wang}
  \email{wang\_b@sjtu.edu.cn}
  
  \address{\textit{Center for Gravitation and Cosmology, College of Physical
      Science and Technology, Yangzhou University, Yangzhou 225009, China}}
  \address{\textit{School of Aeronautics and Astronautics, Shanghai Jiao Tong
      University, Shanghai 200240, China}}
  \begin{abstract}
    
 The phenomenon of spontaneous scalarization of charged black holes has attracted a lot of attention. %From this phenomenon, many theoretical models have been proposed, such as extended Scalar-Tensor-Gauss-Bonnet (eSTGB), Einstein-Maxwell-Scalar (EMS) and Einstein-Maxwell-dilaton (EMD). In these models, studies have shown that various types of scalarized BH solutions are possible, depending on the coupling function that governs the non-minimal coupling between the scalar field and source term. In this phenomenon, the Reissner-Nordstr\"om anti-de Sitter (RN-AdS) BH as a scalar-free BH solution is unstable against scalar perturbations for sufficiently large charge-to-mass ratios. As a result, a new BH configuration with a non-trivial scalar field profile will form as a dynamic evolution endpoint. These scalarized BH solutions bifurcate from scalar-free BH solutions are always energetically favourable. 
In this work, we study the dynamical process of the spontaneous scalarization of charged black hole in asymptotically anti-de Sitter spacetimes in Einstein-Maxwell-scalar models. Including various non-minimal couplings between the scalar field and Maxwell field, we observe that  an  initial scalar-free configuration suffers tachyonic instability and both the scalar field and the black hole irreducible mass grow exponentially at early times and saturate exponentially at late times. For fractional couplings, we find that  though there is negative energy distribution near the black hole horizon, the black hole horizon area never decreases. But when the parameters are large, the evolution endpoints of  linearly unstable bald black holes will be   spacetimes with naked singularity and the cosmic censorship is violated. The effects of the black hole charge, cosmological constant and coupling strength on the dynamical scalarization process are studied in detail. We find that
large enough cosmological constant can prevent the spontaneous scalarization.
 \end{abstract}
 \maketitle
 
 %\tableofcontents
  
  \section{Introduction}
  
Black hole (BH) physics has been an intriguing subject over decades. Recently  high-precision observations have further stimulated interest to study this topic \cite{Berti:2015,Barack:2018}. After the detection of gravitational wave from BH binaries merger \cite{Abbott:2016,Abbott:2017,Abbott:2018} and the observation of BH shadow by Event Horizon Telescope \cite{Cunha:2018,K.Akiyama:2015,K.Akiyama:2017,K.Akiyama:2018}, we have more new windows to disclose deep physics in BHs and examine the validity of general relativity (GR). 
In GR there is a no-hair theorem in BH physics, which indicates that except the mass $M$, charge $Q$ and angular momentum $J$, there is no extra information we can learn from BHs   \cite{Werner:1967,Carter:1971,Chrusciel:2012}. But  the no-hair theorem encounters challenges. Violations were observed in many gravity theories which allow hairy BH solutions, such as those with Yang-Mills field \cite{Volkov:1989,Bizon:1990,Greene:1993,Maeda:1994}, Skyrme field \cite{Luckock:1986,Droz:1991}, conformally-coupled scalar field \cite{Bekenstein:1974} and the dilaton  \cite{Kanti:1996,Zhang:2021,Zhang:2022}. 

In addition to finding new hairy BH solutions to  violate the no-hair theorem, it is of great interest to examine whether there are some relations between  the no-hair BHs and hairy BHs, especially whether there is a mechanism to allow the transition between  them.  Recently   a peculiar dynamical mechanism, the spontaneous scalarization, generating the  hairy BHs has been revived. % \cite{Brihave:2020,Fernandes:2019}.  
This mechanism was first found in the study of neutron stars in scalar-tensor theory  \cite{Damour:1993,Damour:1996,Harada:1997}.  The black holes  can also be spontaneously scalarized if it is surrounded by sufficient amount of matter \cite{Cardoso:20131,Cardoso:2013,Zhang:2014}. The BH spontaneous scalarization is  triggered by the tachyonic instability of the scalar field, through the non-minimal coupling between the  scalar field $\phi$ and a source term $I$. The back-reaction of the scalar instability can destroy the bald BH  and  lead to the formation of a stable scalarized BH which is both thermodynamically and dynamically favored.  %\cite{Fernandes:2019,Fernandes:20191}. 
The source term $I$ can be the Gauss-Bonnet invariant \cite{Doneva:2018,Silva:2018,Antoniou:2018,Blazquez:20180},  the Ricci scalar
for nonconformally invariant black holes  \cite{Herdeiro:2019}, Chern-Simons invariant \cite{Brihaye:2019} or Maxwell  invariant etc.\cite{Herdeiro:2018}.  %All these theoretical achievements have a new cognition of the spontaneous scalarization phenomenon of BHs. 
The study of BH spontaneous scalarization  began in the extended scalar-tensor Gauss-Bonnet (eSTGB) theory  and its  potential relevance in astrophysics has been addressed \cite{Antoniou:2017,Myung:2018,Minamitsuji:2019,Cunha:2019,Macedo:2019sem,Herdeiro:2020wei,Berti:2020kgk,Dima:2020yac,Bakopoulos:2018nui}. However, the equations of motion in the eSTGB theory are difficult to be solved because of the challenging  ill-posedness problem in numerical computations \cite{Ripley:20219,Ripley:2020,East:2020,East:2021}, so that many works limit their dynamical studies in the decoupling limit \cite{Doneva:2021dqn,Kuan:2021lol,Doneva:2021tvn,Silva:2020omi,Doneva:2022byd}.  The  Einstein-Maxwell-scalar (EMS) theory is considered as a technically simpler model,  which has attracted many attentions in examining the dynamics of scalarization, without losing the general interest   \cite{Fernandes:2019,Salcedo:2020,Myung:20190,Astefanesei:2019,Brihave:2020,Fernandes:20191,Xiong:2022ozw}.  It would be fair to say that most available discussions have been concentrated on the static solutions in  asymptotically flat spacetimes. It is of great interest to generalize the discussion to other spacetimes and reveal deeper physics in the dynamical process of scalarization.   % it has been concluded that in flat spacetime,  the scalarized BHs are both thermodynamically and dynamically favoured compared to the bald  BHs. %For different couplings, the eSTGB and EMS models are not always mimic each other.

Considering the special asymptotic boundary in the anti-de Sitter (AdS) spacetime, which behaves as a reflection mirror, it is intriguing to examine whether there are some special properties of the spontaneous scalarization in the AdS spacetime \cite{Guo:2021,Zhang:20211,zhang:20222}. % and what are the differences compared with the situations in the flat spacetime.
On the other hand it is known that the stability of the scalarized BH  depends on the coupling function and the appropriate ranges of parameters in the system \cite{Doneva:2018,Silva:2018,Blazquez:20180}.
In this work, we will carefully investigate the dynamical BH spontaneous scalarization   in EMS theory in AdS spacetimes, and uncover quantitatively the dependence of the dynamical process on the coupling strength between the scalar field and electromagnetic field. 
Furthermore we will reveal the influence of the negative cosmological constant together with other parameters on the dynamical spontaneous scalarization. This can help to have an insight into the special properties of the scalarization in AdS spacetime.

This work is organised as follows. In section 2, we discuss the general framework, introduce the source terms in the  EMS theory, and write out the equations of motion in the Eddington-Finkelstein coordinate. In section 3, we give the conditions  generating spontaneous scalarization, the choices of coupling functions and the boundary conditions of AdS spacetime. The numerical results are presented in section 4. Finally, we summarize and discuss the results obtained. 

 \section{Model setup}
 \subsection{The Action and Equations of Motion}
 The action we consider in this work is 
 \begin{equation}
   S= -\frac{1}{16\pi}\int d^{4}x\sqrt{-g}[R-2\Lambda-2\partial{_\mu}\phi\partial{^\mu}\phi-f_{i}(\phi)I(\psi;g) ].\label{eq:a}
 \end{equation}
Here $R$ is the Ricci scalar, $\Lambda=-3/L^{2}$ is the cosmological constant with the AdS radius $L$. The scalar field $\phi$ is minimally coupled to the metric $g_{\mu\nu}$ and non-minimally coupled to the source  term $I(\psi,g)$ which  generically depends on the spacetime metric $g_{\mu\nu}$ and the extra matter fields, collectively denoted by $\psi$. The subscript   $i$ in coupling function $f_{i}(\phi)$  will be used to label the various coupling choices. %, as specified below.
In EMS theory the extra matter field is a gauge field $A_{\mu}$ with  
 \begin{equation}
   I(\psi,g)=F_{\mu\nu}F^{\mu\nu}
 \end{equation}
in which $F_{\mu\nu}=\partial_{\mu}A_{\nu}-\partial_{\nu}A_{\mu}$ is electromagnetic field strength tensor. In eSTGB theory, the source term is Gauss-Bonnet invariant  $I(\psi;g)=R^2-4R_{\mu\nu}R^{\mu\nu}+R_{\mu\nu\rho\sigma}R^{\mu\nu\rho\sigma}$ and $\psi=0$, $i.e.$ without any extra material fields.

The field equations obtained by varying the   action  with respect to the   $g_{\mu\nu}$, $\phi$ and $A_{\mu}$ are
\begin{align}
  R_{\mu\nu}-\frac{1}{2}Rg_{\mu\nu}+\Lambda g_{\mu\nu} ={}& 2\Bigl[\partial_{\mu}\phi\partial_{\nu}\phi-\frac{1}{2}g_{\mu\nu}\partial_{\rho}\phi\partial^{\rho}\phi + f(\phi) (F_{\mu\rho}{F_{\nu}}^{\rho}-\frac{1}{4}g_{\mu\nu}F_{\rho\sigma}F^{\rho\sigma})\Bigr].\label{eq:b}\\
  \frac{1}{\sqrt{-g}}\partial_{\mu}\left(\sqrt{-g}\partial^{\mu}\phi\right)={}& \frac{1}{4}\frac{df(\phi)}{d\phi}F_{\rho\sigma}F^{\rho\sigma}.\label{eq:c}\\
  \partial_{\mu}\left(\sqrt{-g}f(\phi)F^{\mu\nu}\right) ={}& 0.\label{eq:d}
\end{align}

\subsection{Conditions for Spontaneous Scalarization of Black Holes}
We assume that the model admits scalar-free solutions, i.e., $\phi=0$ satisfies the equations of motion (\ref{eq:b})-(\ref{eq:d}). The coupling function $f(\phi)$ must obey the following criteria:

 1) $f(\phi)\mid_{\phi=0,r\to\infty}=1$. The system    approaches the electromagnetic vacuum in the far region.%, and in eSTGB model, $f(\phi)\mid_{\phi=0,r\to\infty}=0$;
 
2) ${\frac{df(\phi)}{d\phi}}\mid_{\phi=0}=0$. This allows the existence of scalar-free solution;
 
 3) $\frac{d^2f(\phi)}{d\phi^2}\mid_{\phi=0}>0$. This  guarantees the appearance of the tachyonic instability which drives the system away from the scalar-free solution.%, whereas in eSTGB it is reversed.
 
 In fact, to guarantee the existence of non-trivial scalarized BHs, one can also derive the constraints equivalent to conditions 1) and 3) from eq.~(\ref{eq:c}) in the case of purely electric (or magnetic) RN BHs, which is the so-called Bekenstein--type inequality 
$f(\phi)_{,\phi\phi}>0$ and $\phi{f_{,\phi}>0}$  \cite{Astefanesei:2019}.
 
 \subsection{Selection of Coupling Function}
 In this work we   simulate the dynamical evolution of the BH spontaneous scalarization in EMS theory in AdS spacetime with coupling functions satisfying the above conditions, which include 
 
 $i){}$: a fractional coupling  $f_{F}(\phi)=\frac{1}{1+b\phi^{2}}$;
 
 $ii){}$: a hyperbolic cossine coupling  $f_{H}(\phi)=\cosh(\sqrt{-2b}\phi)$;
 
 $iii){}$: a power coupling $f_{P}(\phi)=1-b\phi^{2}$.
 
The parameter $b$ is a dimensionless constant in all cases. Note that they have the same leading order expansion for small $\phi$.

\section{Numerical Setup}
\subsection{Equations of Motion in Eddington-Finkelstein Coordinate}
We study  the  dynamical  formation of a  charged scalarized BHs  from a spherically symmetric scalar-free RN-AdS BH  suffering tachyonic instability in EMS theory, by adopting the ingoing Eddington-Finkelstein coordinate ansatz
\begin{equation}
  ds^{2} = -\alpha(t,r)dt^{2} + 2dtdr + \zeta(t,r)^{2} \left(d\theta^{2}+\sin^{2}\theta d\varphi^{2}).\right.\label{eq:e}
\end{equation}
Here $\alpha(t,r)$ and $\zeta(t,r)$ are the metric functions. They are regular on the BH apparent horizon which satisfies $  g^{\mu\nu}\partial_\mu{\zeta}\partial_\nu{\zeta}=0.$
We choose the gauge field  as
\begin{equation}
  A_{\mu}dx^{\mu}=A\left(t,r\right)dt.  \label{eq:g}
\end{equation}
Plugging the above ansatz into    (\ref{eq:d})   yields the first integral
\begin{equation}
  \partial_{r}A=\frac{Q}{\zeta^{2}f\left(\phi\right)},
\end{equation}
in which Q is an integral constant interpreted as the electric charge.
To implement the numerical method, we introduce auxiliary variables
\begin{align}
  S&=\partial_{t}\zeta+\frac{1}{2}\alpha\partial_{r}\zeta.\label{eq:h}\\ P&=\partial_{t}\phi+\frac{1}{2}\alpha\partial_{r}\phi.\label{eq:i}
\end{align}
Substituting  these   into   (\ref{eq:b}), we get
\begin{align}
  \partial_{t}S ={}& \frac{1}{2}S\partial_{r}\alpha+\frac{\alpha}{2}\left(\frac{2S\partial_{r}\zeta-1}{2\zeta}+\frac{1}{2}\zeta\Lambda+\frac{Q^2}{2\zeta^{3}f(\phi)}\right)-\zeta P^2.\label{eq:j}\\
  \partial^{2}_{r}\alpha={}&-4P\partial_{r}\phi+\frac{4S\partial_{r}\zeta-2}{\zeta^2}+\frac{4Q^2}{\zeta^4f(\phi)}.\label{eq:k}\\
  \partial_{r}S={}&\frac{1-2S\partial_{r}\zeta}{2\zeta}-\frac{\zeta\Lambda}{2}-\frac{Q^2}{2\zeta^3f(\phi)}.\label{eq:l}\\
  \partial^{2}_{r}\zeta={}&-\zeta\left(\partial_{r}\phi\right)^2.\label{eq:m}
\end{align}
The scalar equation (\ref{eq:c}) gives
\begin{equation}
  \partial_{r}P=-\frac{P\partial_{r}\zeta+S\partial_{r}\phi}{\zeta}-\frac{Q^2}{4\zeta^4f(\phi)^2}\frac{df(\phi)}{d\phi}.\label{eq:n}
\end{equation}
As long as the initial   $\phi$ is given, we can integrate constraint equations (\ref{eq:k})-(\ref{eq:n}) to get initial $\alpha,S,\zeta,P$. The $\phi$ on the next time slice can be obtained from the evolution equation (\ref{eq:i}). This formulation has been widely used to simulate the
nonlinear dynamics in AdS spacetimes due to its simplicity
and high accuracy  \cite{Zhang:20211,zhang:20222,Chesler:2009,Bhaseen:2013,Chesler:2013,Janik:2017,Chesler:2019,Bosch:2016,Bosch:2019}. 

\subsection{Boundary Conditions of AdS Spacetime}
 To solve the set of differential equations numerically, we have to implement suitable boundary conditions. % for the desired functions and corresponding derivatives. 
 An asymptotic approximation of the variables in the far region takes the form
 \begin{align}
  \phi={}&\frac{\phi_{3}(t)}{r^{3}}+\frac{3}{8\Lambda r^{4}}\left(-bQ^2-8\phi'_{3}(t)\right)+O\left(r^{-5}\right).\label{eq:o}\\
  \alpha={}&-\frac{\Lambda}{3}r^{2}+1-\frac{2M}{r}+\frac{Q^2}{r^2}+\frac{\Lambda}{5r^{4}}\phi^{2}_{3}(t)+O\left(r^{-5}\right).\label{eq:p}\\
  \zeta={}&r-\frac{3\phi^{2}_{3}(t)}{10r^{5}}+\frac{3\phi_{3}(t)}{14\Lambda r^{6}}\left(-bQ^{2}-8\phi'_3(t)\right)+O\left(r^{-7}\right).\label{eq:q}\\
  S={}&-\frac{\Lambda}{6}r^{2}+\frac{1}{2}-\frac{M}{r}+\frac{Q^{2}}{2r^{2}}-\frac{3\Lambda}{20r^{4}}\phi^{2}_{3}(t)+O\left(r^{-5}\right).\label{eq:r}\\
  P={}&\frac{\Lambda\phi_{3}(t)}{2r^2}+\frac{1}{r^{3}}\left(\frac{-bQ^2}{4}-\phi'_{3}(t)\right)+\frac{3}{2\Lambda r^{4}}\phi''_{3}(t)+O\left(r^{-5}\right).\label{eq:s}
 \end{align}
 in which $\phi'_{3}(t)=\frac{d\phi_{3}(t)}{dt}$. This series expansion contains three constants: the ADM mass $M$, the charge $Q$ of BH, and the cosmological constant $\Lambda$. Hereafter, we fix  the value of ADM mass as $M=1$ in this work to implement the dimensionless of the physical quantities. Meanwhile, we study the BH irreducible mass $M_{ir}$ and the rescaled Misner-Sharp mass $M_{ms}$, which are respectively defined as
 \begin{align}
  M_{ir}&=\sqrt{\frac{A_H}{4\pi}}=\zeta\left(t,r_H\right),\label{eq:t}\\
  M_{ms}&=\frac{m}{4\pi}=\frac{1}{2}\zeta\left(1-\frac{\Lambda}{3}\zeta^{2}-g^{\mu\nu}\partial_\mu{\zeta}\partial_\nu{\zeta}\right).\label{eq:u}
 \end{align}
 Here $A_H=4\pi\zeta^2\left(t,r_H\right)$ and $r_H$ stands for the coordinate location of the BH apparent horizon. The irreducible mass equals the horizon area radius. At static case, $\phi_{3}$ can be viewed as the scalar charge indicating the existence of the scalar hair. But it is unknown here and needs to determined by evolution. Notice that some of the variables in the series expansion above like $\alpha,\zeta,S$ are divergent at infinity. Therefore, the following new variables are introduced for numerical calculation. 
 \begin{equation}
   \begin{aligned}
     \zeta\equiv r\sigma,\quad\alpha\equiv r^{2}a,\quad S\equiv r^{2}s,\quad P\equiv\frac{1}{r}p.
   \end{aligned}
 \end{equation}
 In addition, the scalar perturbation in AdS spacetime can reach the spacial infinity at finite coordinate time and be bounced back to the bulk. So  the spacial infinity must be included in the computational domain. The effective way is to compactify the radial direction by a coordinate transformation $i.e.~z=\frac{r}{r+M}$. In this new coordinates, the computational domain that we take as $(z_i,1)$, and where $z_i$ is close but smaller than the initial BH apparent horizon and $z=1$ corresponds to spacial infinity. From the above conditions, we can obtain that the boundary conditions at infinity:
 \begin{equation}
  \begin{aligned}
    \sigma=1,\sigma'=0,s=&-\frac{\Lambda}{6},s'=0,s''=6\left(M-1\right),p=0,\\a=&-\frac{\Lambda}{3},a'=0,a''=12\left(M-1\right).
  \end{aligned}
 \end{equation}
 Here the prime denotes the derivative with respect to $z$. 
 For the initial profiles of the scalar field, we take the Gaussian wave packet
 \begin{equation}
  \phi_{0}=ae^{-\left({\frac{r-cM}{wM}}\right)^{2}}.\label{eq:w}
 \end{equation}
 Here $a,c,w$ parameterize the initial amplitude, center and width of the Gaussian wave, respectively. %We employ the same method as in reference \cite{Zhang:2021,Zhang:20211} to implement the numerical evolution. %Substantially, let us start by exhibiting some typical solutions obtained from the numerical integration.
 
 \section{Numerical Results}
 \subsection{Results for Fractional Coupling}
 \subsubsection{Scalar field for fractional coupling $f_{F}(\phi)$}

\begin{figure}[htbp]
   \begin{minipage}[t]{0.3\linewidth}
     \centering
     \includegraphics[width=4.8cm,height=4.5cm]{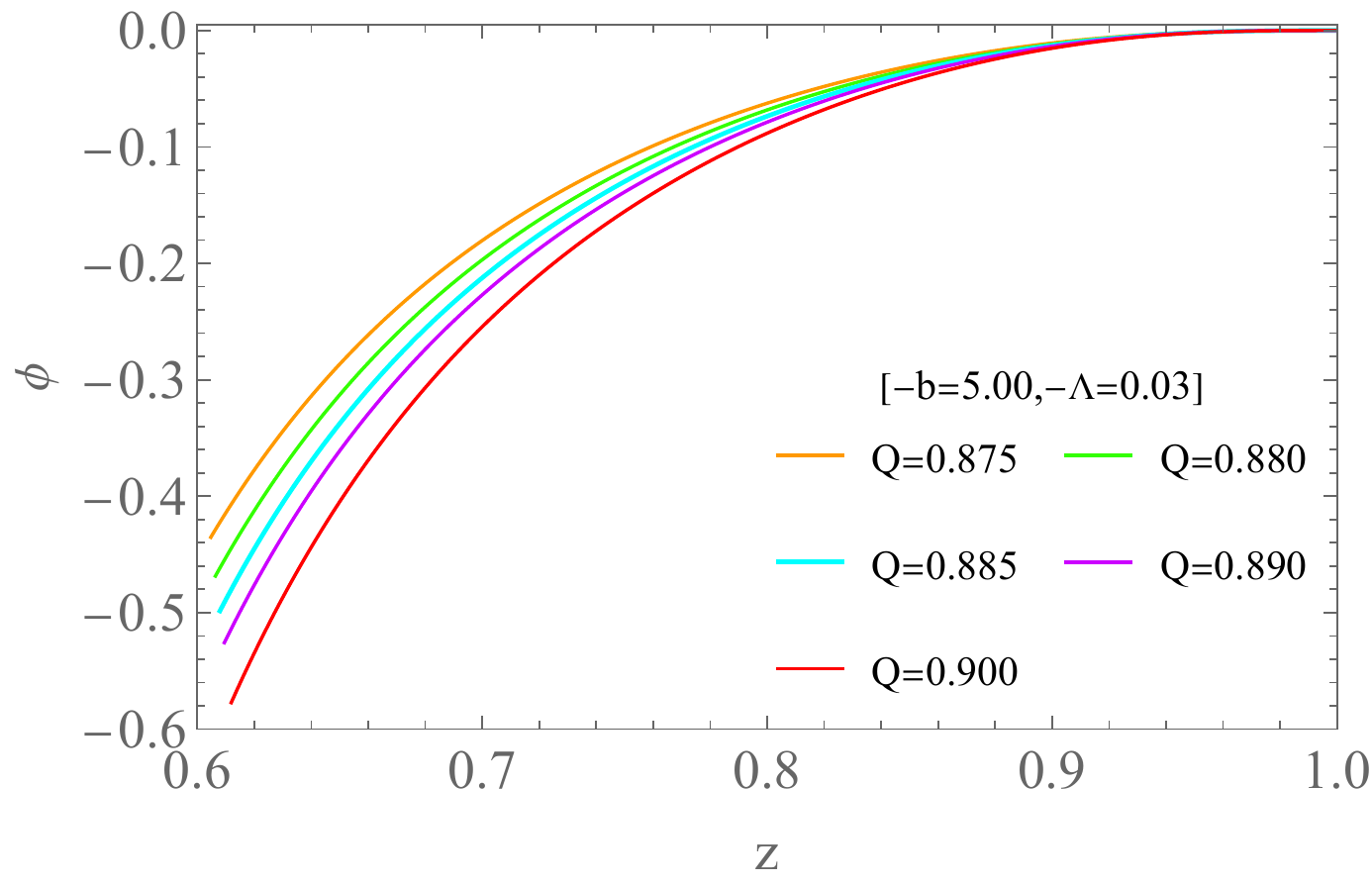}
   \end{minipage}
   \begin{minipage}[t]{0.3\linewidth}
     \centering     
     \includegraphics[width=4.8cm,height=4.5cm]{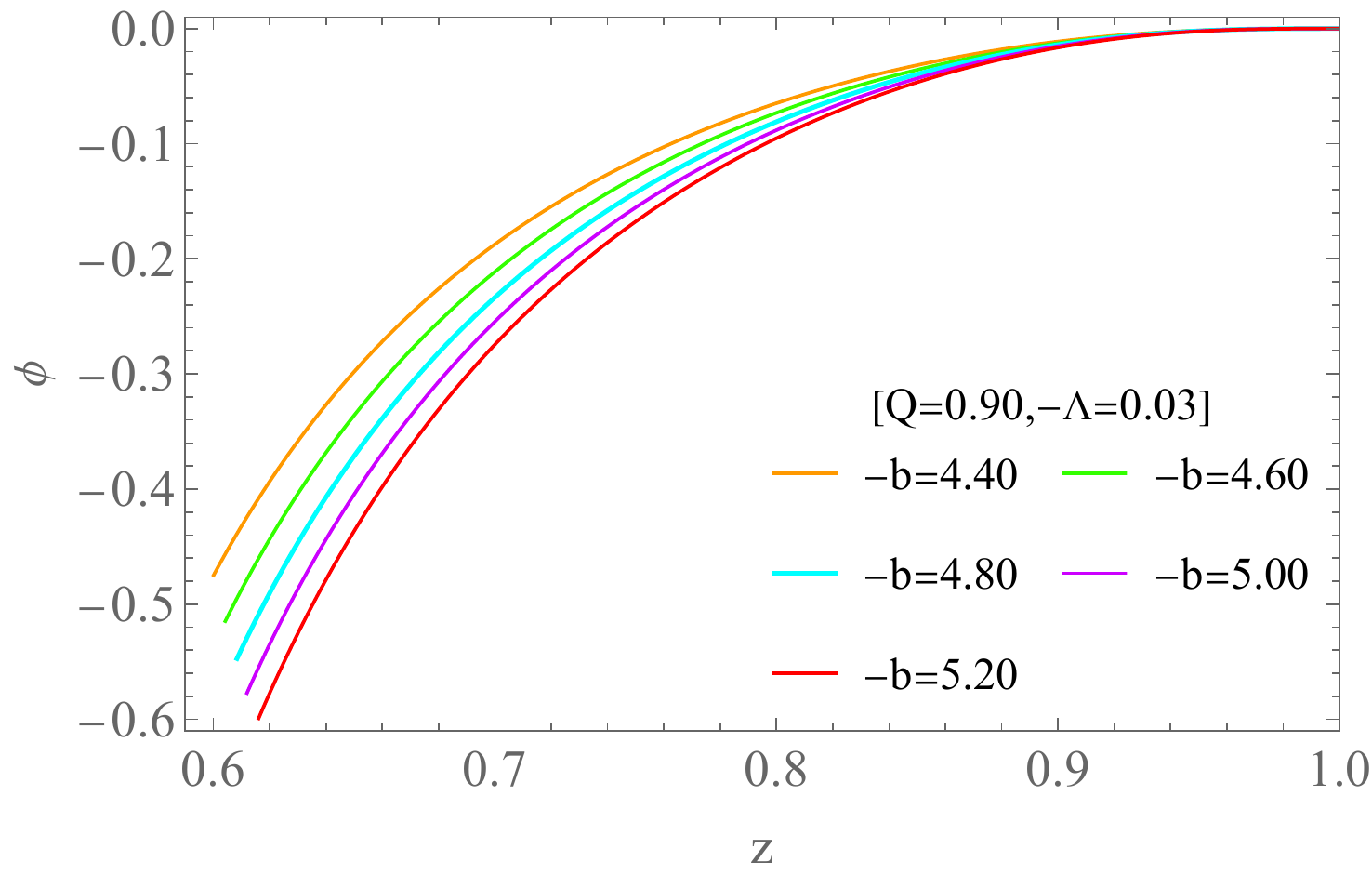}
   \end{minipage}
   \begin{minipage}[t]{0.3\linewidth}
     \centering
     \includegraphics[width=4.8cm,height=4.5cm]{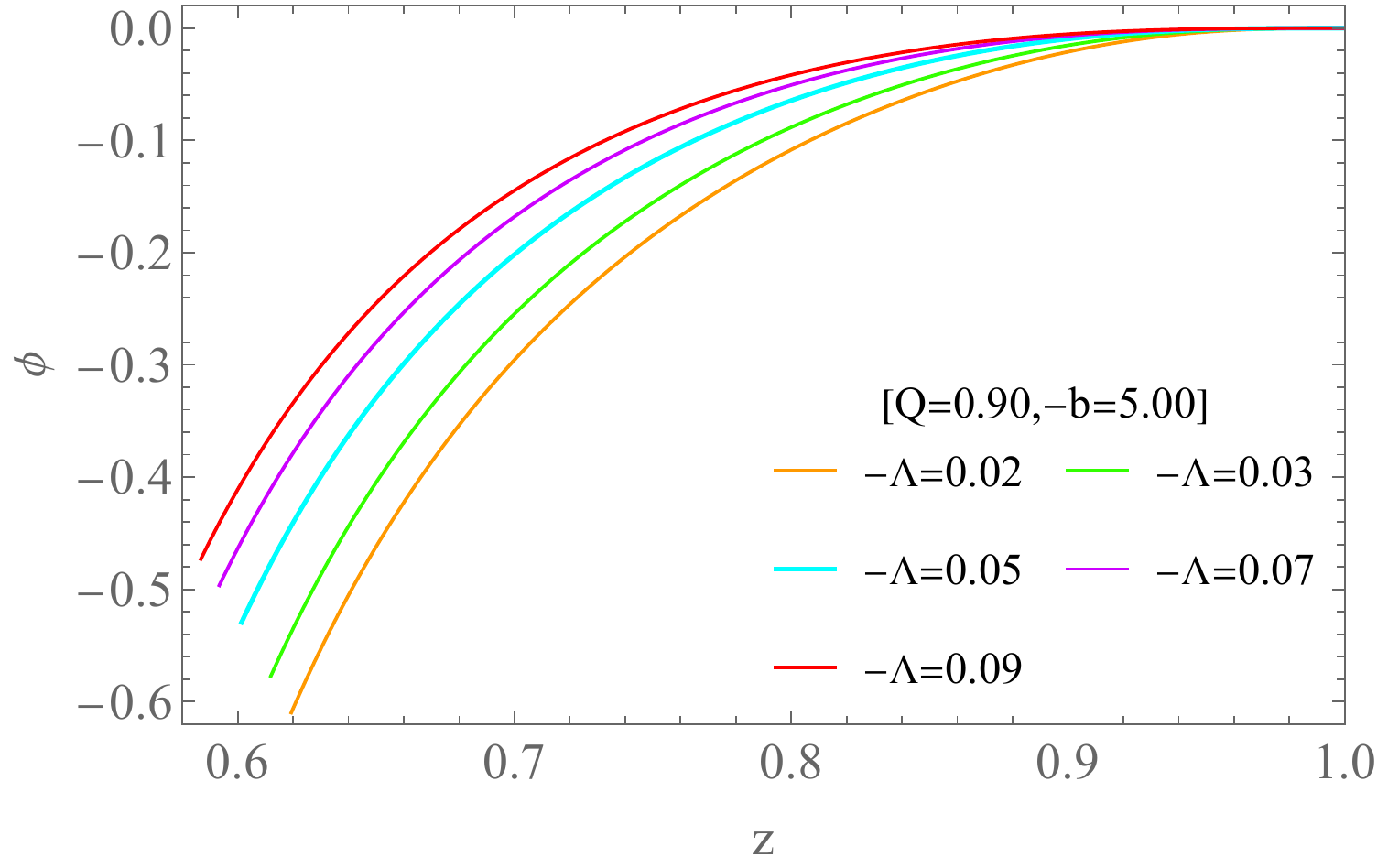}
   \end{minipage}
 {\footnotesize{}\caption{{\footnotesize{}\label{fig:1}The spatial distribution of the scalar field $\phi$ outside the horizon for various charge $Q$, coupling constant $b$ and cosmological constant $\Lambda$ when the system reaches equilibrium. % are all asymptotically vanishing after the dynamic evolution of the system is stabilized. A common feature of this nodeless solution is that the extrema all pile up at the horizon (denoted as $\phi_{H}$) and then disappear asymptotically at radial infinity.
}}}{\footnotesize\par}
 \end{figure}

We first investigate the final spatial distribution of the scalar field when the system reaches equilibrium starting from  an unstable  RN-AdS BH with fractional coupling function under initial scalar perturbation. As shown in Fig.~\ref{fig:1}, an obvious feature is that the scalar field piles up at the horizon. It is  nodeless and monotonically tends to zero in all situations. The final scalar field value on the BH horizon  grows with  $Q$ and $-b$, while decreases with $\Lambda$.
 %and move away from the center of the BH. The reason for this phenomenon is that when the values of $Q$ and $-b$ are larger, the repulsion effect triggered by the effective tachyon mass is stronger and more energy will be transferred from the Maxwell field to the scalar field. As a result, the repulsion effect can effectively repel more scalar field outside the horizon, though, the BH can still capture more scalar field so that the horizon expands. 
 \begin{figure}[htbp]
   \begin{minipage}[t]{0.3\linewidth}
     \centering
     \includegraphics[width=4.8cm,height=9cm]{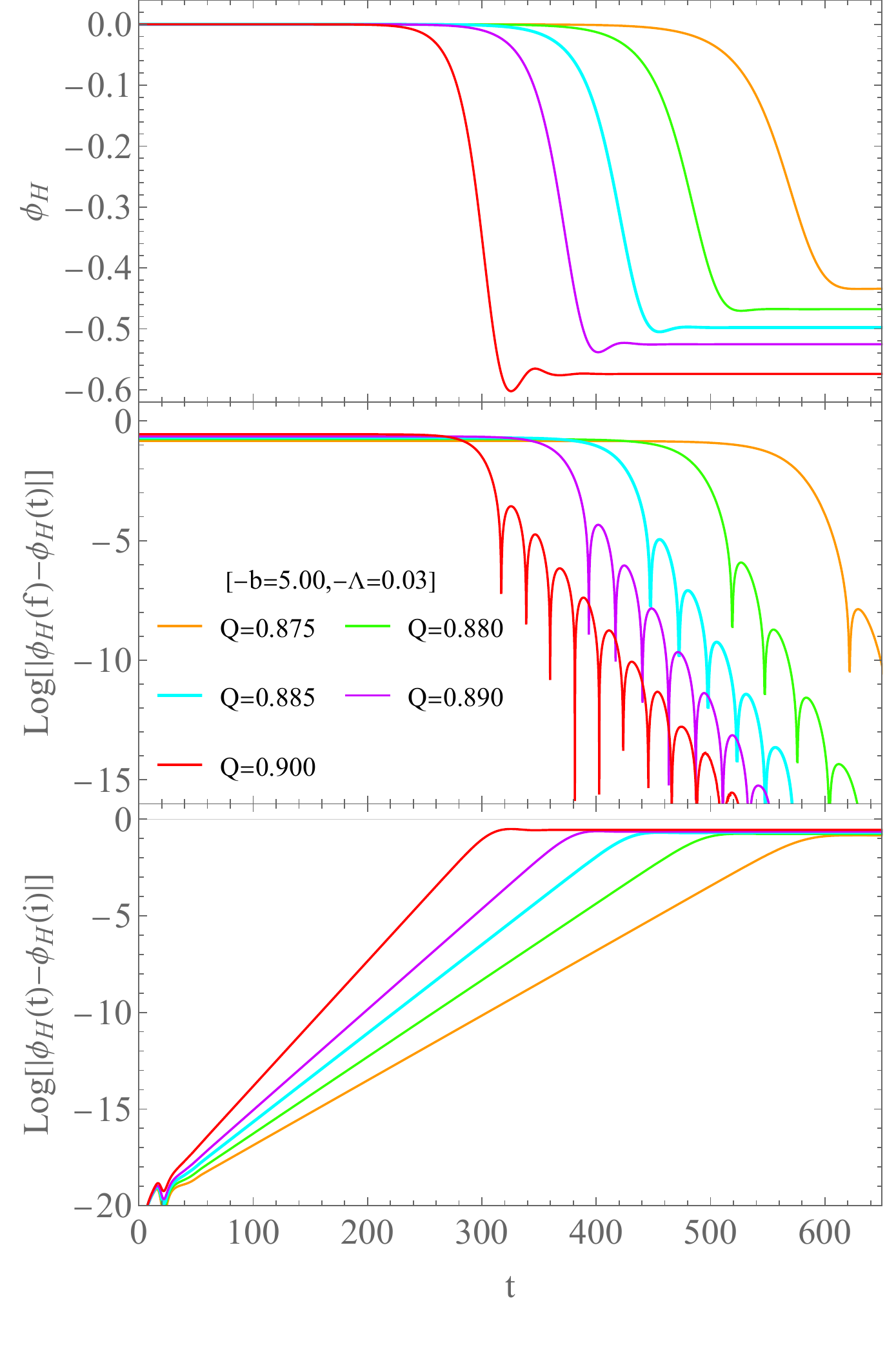}
   \end{minipage}
   \begin{minipage}[t]{0.3\linewidth}
     \centering     
     \includegraphics[width=4.8cm,height=9cm]{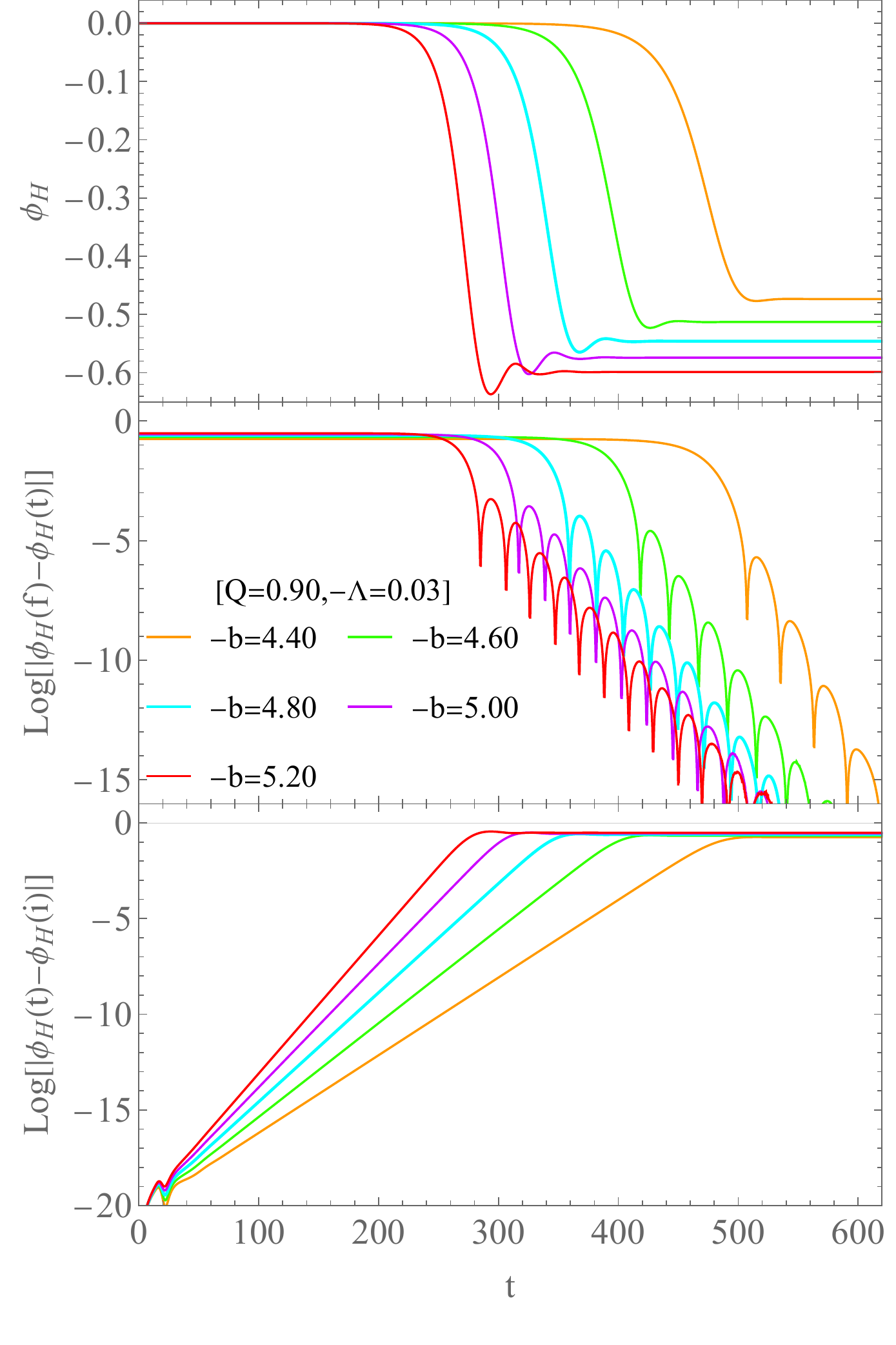}
   \end{minipage}
   \begin{minipage}[t]{0.3\linewidth}
     \centering
     \includegraphics[width=4.8cm,height=9cm]{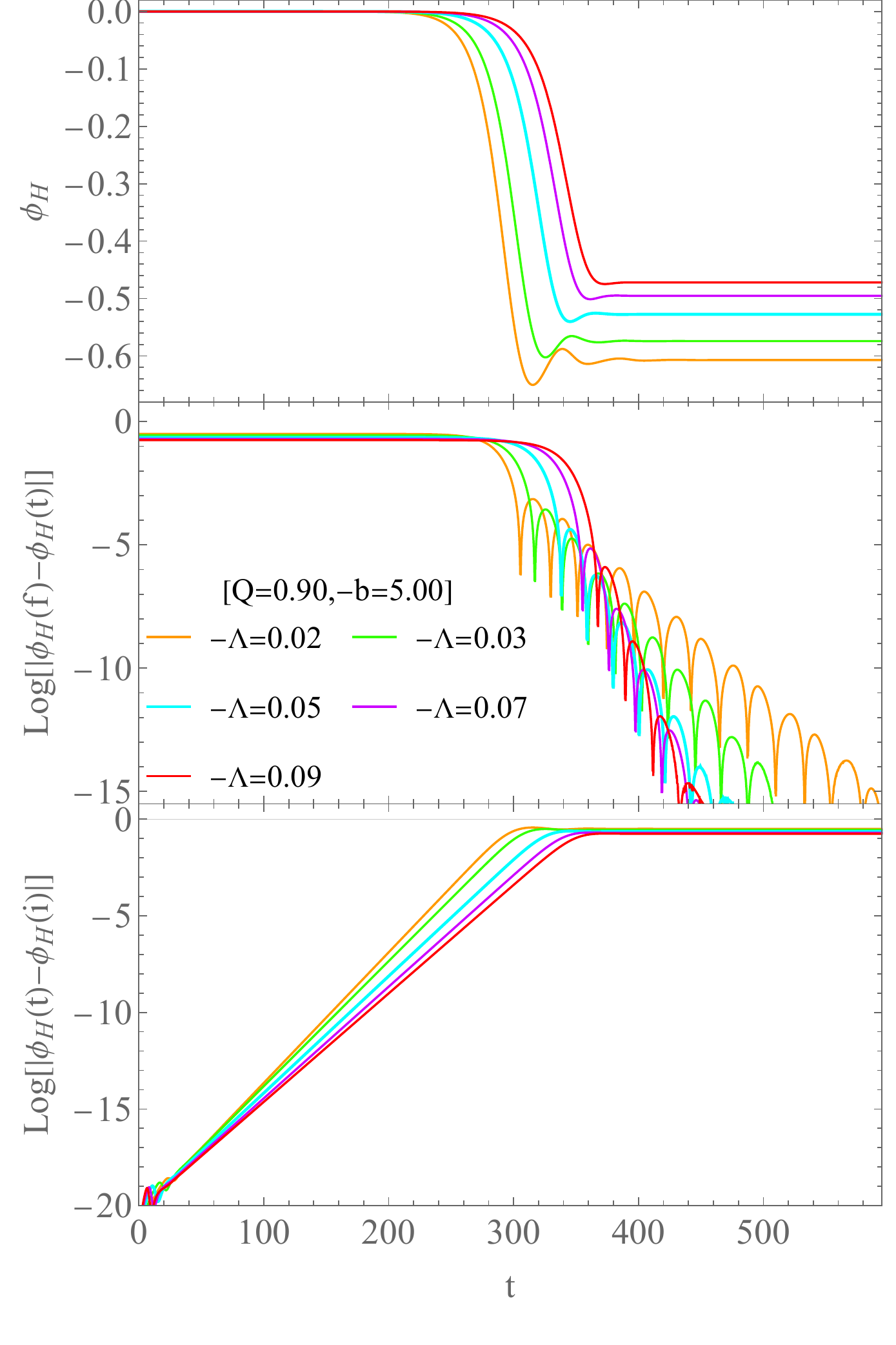}
   \end{minipage}
 {\footnotesize{}\caption{{\footnotesize{}\label{fig:2}The upper row shows the evolution of the scalar field value $\phi_{H}$ on the horizon. The  lower and center rows indicate that $\phi_{H}$  grows exponentially at first and then saturates to an equilibrium value with damped oscillation.%, where $\phi_{H}(f)$ represents the scalar field after evolution equilibrium at the horizon.
}}}{\footnotesize\par}
 \end{figure}
 
 To figure out how the system evolves from the initial bald RN-AdS BH to the final hairy BH, we show the evolution of the scalar field value on the horizon  $\phi_{H}$ in the upper row of Fig.~\ref{fig:2}. One can find that the BH is decorated with scalar hair faster and more heavily for larger $Q$ and stronger coupling $-b$ between the scalar field and Maxwell field. On the contrary, the cosmological constant $\Lambda$ suppresses this phenomenon. These are consistent with the results from Fig.~\ref{fig:1}. 

 In the middle and lower rows of Fig.~\ref{fig:2}, we show the evolution of  $\log|\phi_{H}(f)-\phi_{H}(t)|$  and $\log|\phi_{H}(t)-\phi_{H}(i)|$. Here $\phi_H(i)=0$ and $\phi_H(f)$ are the initial and final scalar field value on the horizon, respectively. The lower row implies that if the RN-AdS BH is in the unstable regime, any initial arbitrarily small perturbation will result in an exponential growth of the scalar field at early times. The middle row implies that the scalar field saturates to an equilibrium value at late times and the final equilibrium BH is endowed with scalar hair. Hence the evolution of the scalar field on the horizon  can be approximated by 
 \begin{equation} 
 \phi_{H}\approx\begin{cases}
   %\phi_{H}(i)+
   \exp(\nu_i{t}+\nu_1), & \text{early times,}\\
   \phi_{H}(f)-\exp(-\nu_f{t}+\nu_2), & \text{late times.}
  \end{cases} 
 \end{equation} 
Here $\nu_i$ is the growth rate of $\phi_{H}$  at early times and $\nu_f$ is the imaginary part of dominant mode frequency at late times.  $\nu_{1,2}$ are some subdominant terms depending on $Q$, $b$ and $\Lambda$. The lower row of Fig.~\ref{fig:2} reveals that $\nu_i$ is positively related to $Q$ and $-b$, and negatively related to $-\Lambda$, which means that the time of scalarized BH bifurcating from the initial RN-AdS BH will be shortened during the growth stage for larger $Q$ and $-b$, and prolonged for larger  $-\Lambda$. At late times, however, the central row of Fig.~\ref{fig:2} shows that during the saturation stage, $\phi_{H}$ takes 
longer time to converge to its final value for larger $Q,-b$ and smaller $-\Lambda$. On the other hand, the relations between $\nu_f$ and $Q,b,\Lambda$ are contrary to those of $\nu_i$.

 \subsubsection{Misner-Sharp mass of fractional coupling}
 
\begin{figure}[htbp]
    \begin{minipage}{0.3\linewidth}
      \centering
      \includegraphics[width=4.8cm,height=8cm]{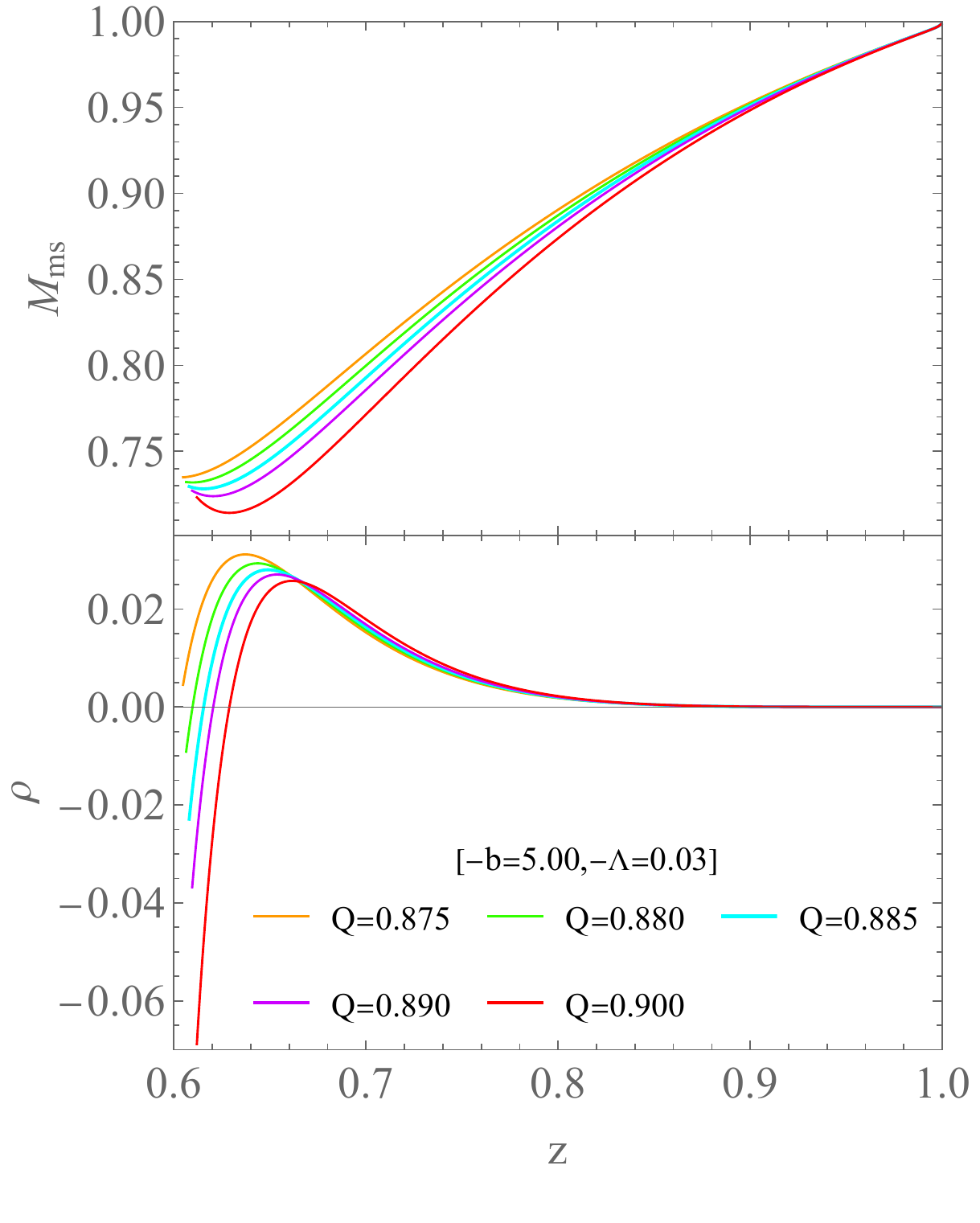}
    \end{minipage}
    \begin{minipage}{0.3\linewidth}
      \centering
      \includegraphics[width=4.8cm,height=8cm]{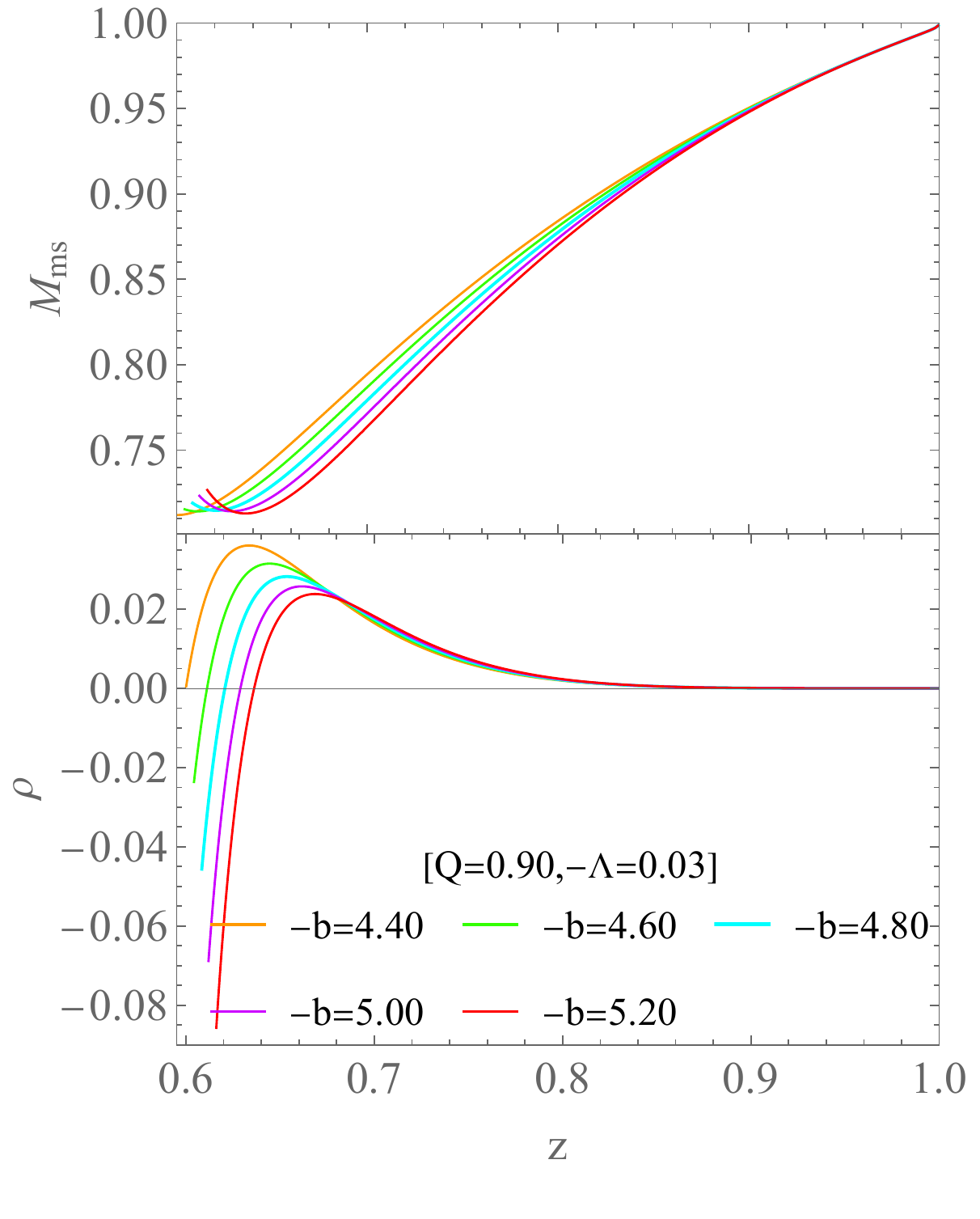}
    \end{minipage}
    \begin{minipage}{0.3\linewidth}
      \centering
      \includegraphics[width=4.8cm,height=8cm]{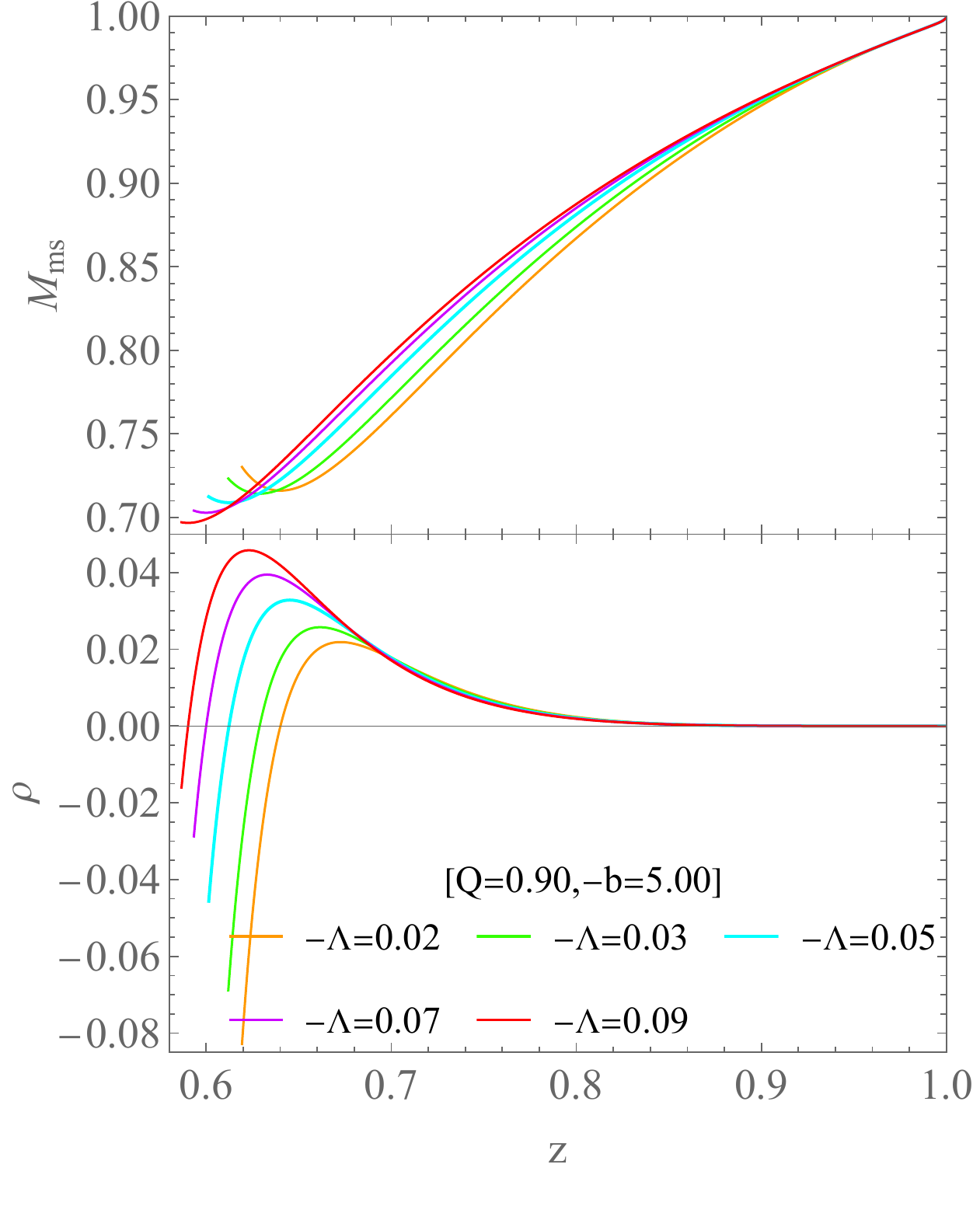}
    \end{minipage}
    {\footnotesize{}\caption{{\footnotesize{}\label{fig:4}Fractional  coupled scalarized BH solutions exhibit negative energy densities $\rho$ in the vicinity of horizon. Note that the left endpoints locate  on the BH horizon and we only show the distribution outside the  horizon.
    }}
    }{\footnotesize\par}
  \end{figure}
The Misner-Sharp mass $M_{ms}$ of scalarized solutions is a function of radius and time. Its final distribution when the system reaches equilibrium is exhibited in the upper row of Fig.~\ref{fig:4}. It increases to the ADM mass $M=1$ as the radius tends to infinity. However,  in the near horizon region,  the $M_{ms}$ decreases with radius for large $Q,-b$ and small $-\Lambda$. This implies that there are negative energy    distribution near the black hole. 
In fact, for static solution, the energy density can be expressed as
 \begin{equation}
   \rho=\frac{\alpha}{2}\left(\frac{\partial{\phi}}{\partial{r}}\right)^{2}+\frac{Q^{2}(1+b\phi^2)}{2\zeta^{4}}, \label{eq:rho}
 \end{equation}
 which follows from
  $\rho=T_{\mu\nu}Z^{\mu}Z^{\nu}$. Here $T_{\mu\nu}$ is the stress energy tensor in eq.~(\ref{eq:b}), and $Z^\mu=(\partial_t)^\mu/\sqrt{\alpha}$.
The energy density distribution is shown in the lower row of Fig.~\ref{fig:4}. One can find that the scalarized BH solution obtained with fractional coupling does have negative energy density in the vicinity of horizon.  This is similar to the results found in asymptotically flat spacetime \cite{Fernandes:2019}.
Actually,  the negative energy originates from the second term in   eq.~(\ref{eq:rho}) since $1+b\phi^{2}<0$ in the vicinity of the horizon. The negative contribution is more significant for stronger coupling and larger charge. 

 \begin{figure}[htbp]
    \begin{minipage}{0.3\linewidth}
      \centering
      \includegraphics[width=4.8cm,height=4.5cm]{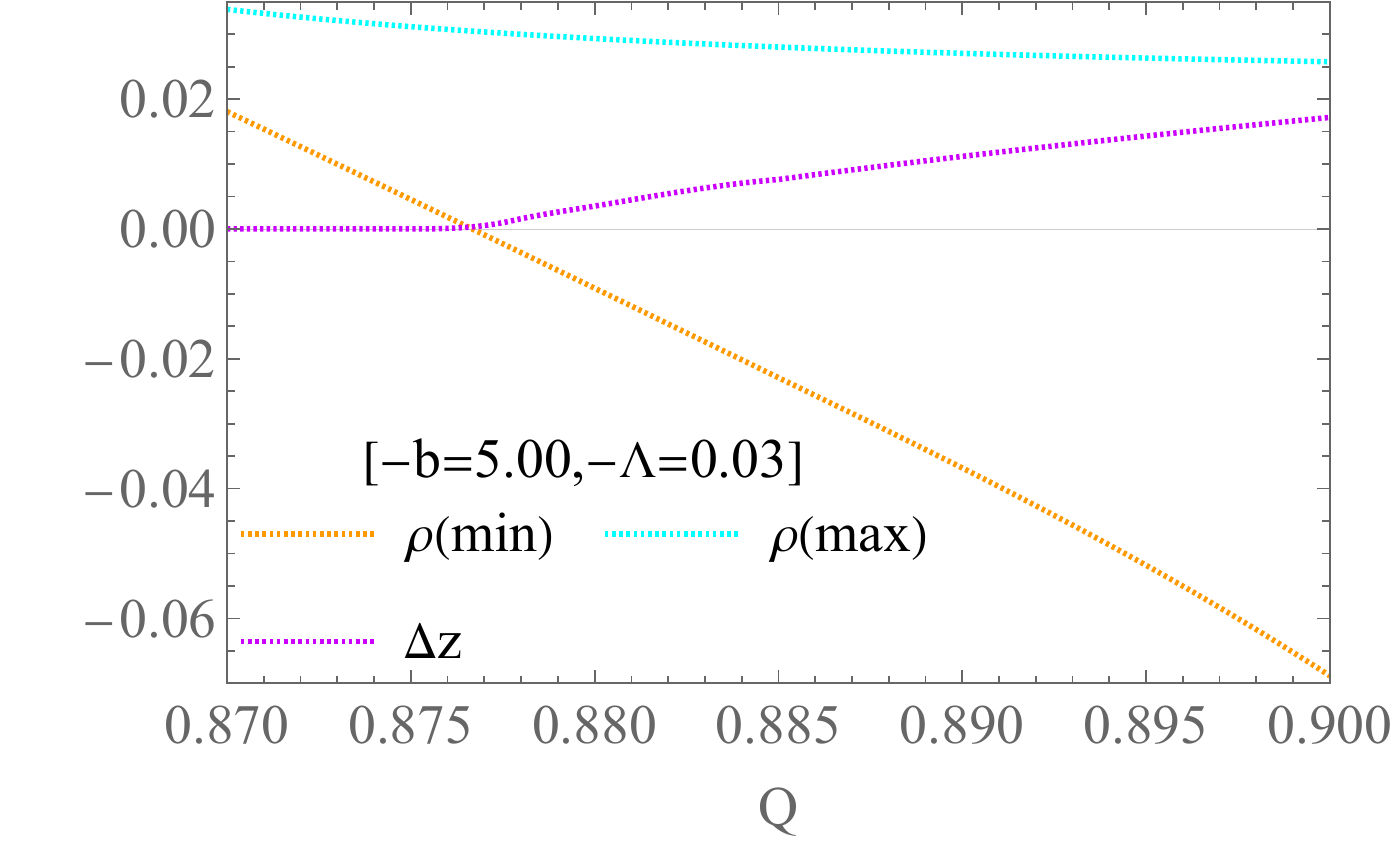}
    \end{minipage}
    \begin{minipage}{0.3\linewidth}
      \centering
      \includegraphics[width=4.8cm,height=4.5cm]{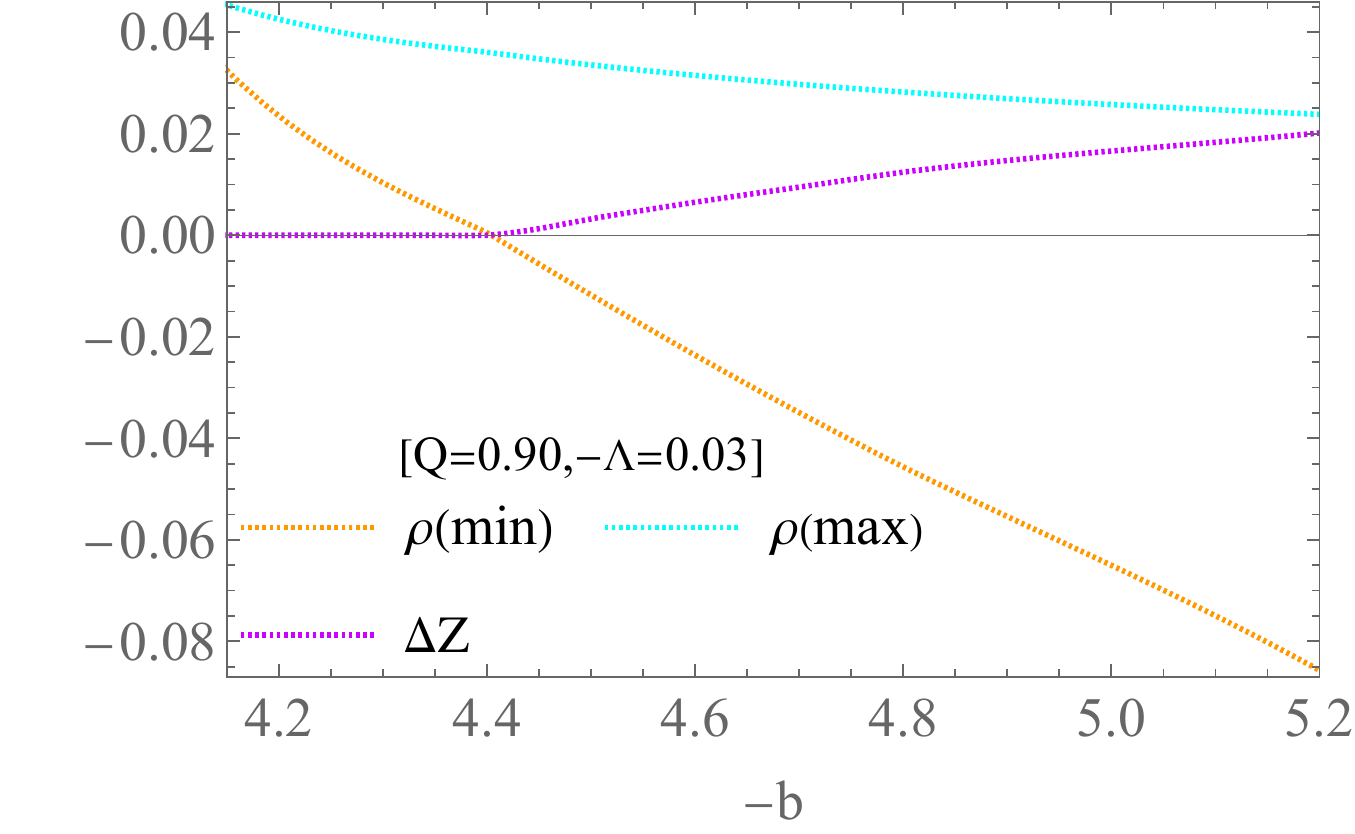}
    \end{minipage}
    \begin{minipage}{0.3\linewidth}
      \centering
      \includegraphics[width=4.8cm,height=4.5cm]{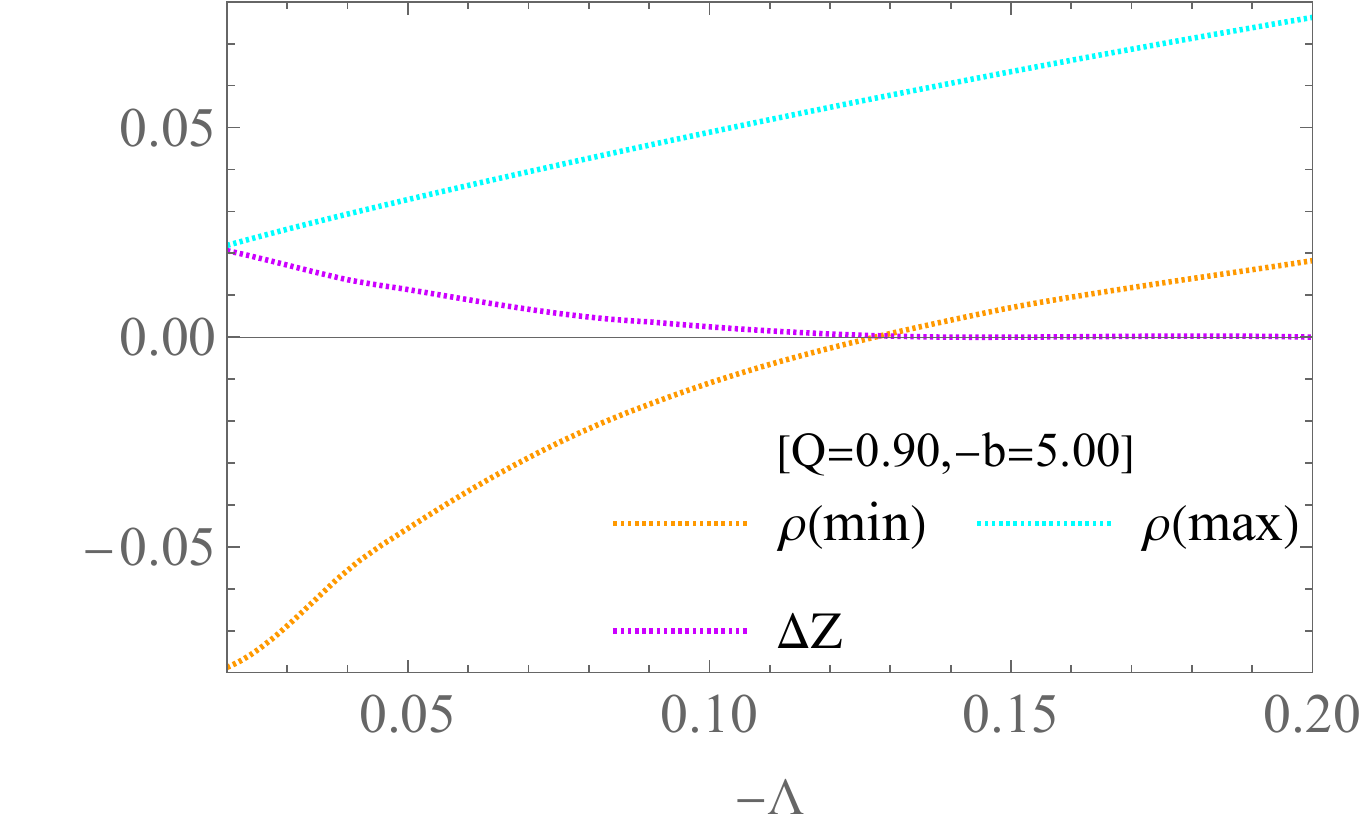}
    \end{minipage}
    {\footnotesize{}\caption{{\footnotesize{}\label{fig:5}The  maximum and minimum of the energy density $\rho$ outside the horizon, the negative energy band $\Delta{z}$ and the minimum of the Misner-Sharp mass versus $Q,b$ and  $\Lambda$.}}
    }{\footnotesize\par}
  \end{figure}
  
The extremum of $\rho$ and the negative energy band $\Delta{z}$ are shown in Fig.~\ref{fig:5}. The   $\rho(min)$ decreases monotonically with  $Q$ or $-b$, while $\Delta{z}$ first remains zero and then increases. %\red{, which is different from flat spacetime \cite{Fernandes:2019}.}
This result can also be explained by eq.~(\ref{eq:rho}) in which the first term  is always positive outside the horizon. For small $Q$ or $-b$, the final scalarized BH has less hair so that the first term  is larger than the second term, so the energy density $\rho$ is positive and the negative energy band $\Delta{z}$ is zero. 
%Compared to $Q$ and $b$, however, research shows that $\Lambda$ hates the emergence of rule-violating solutions. 
On the one hand, the right row of Fig.~\ref{fig:5} shows that the increase of $-\Lambda$ suppresses the negative energy distribution outside the horizon. %, and on the other hand, also pushes the negative energy density that violates the energy condition into the horizon. Thus, one can see that $M_{ms}$ is continuously decreasing and approaching the horizon as $-\Lambda$ increases but the behavior in $Q$ and $-b$ is reversed. As for the peak value $\rho(max)$ of $\rho$, it reflects that $M_{ms}$ has a common feature among these free parameters, that is, $M_{ms}$ can quickly converge to the ADM mass compared with the situation where the negative energy band $\Delta{z}$ is wider.

\subsubsection{Naked singularity}

 \begin{figure}[htbp]
    \begin{minipage}{0.45\linewidth}
      \centering
      \includegraphics[width=6.8cm,height=5.5cm]{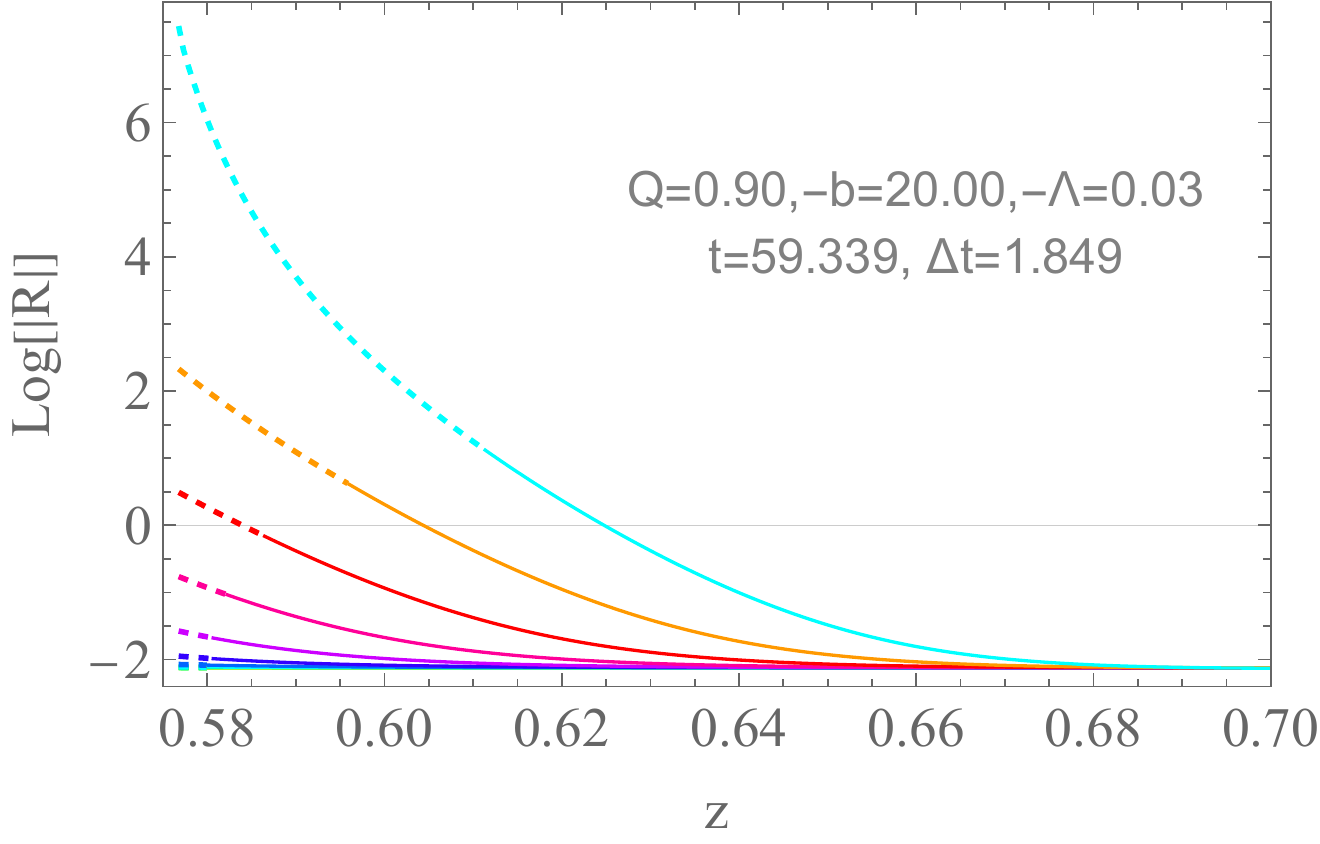}
    \end{minipage}
    \begin{minipage}{0.45\linewidth}
      \centering
      \includegraphics[width=6.8cm,height=5.5cm]{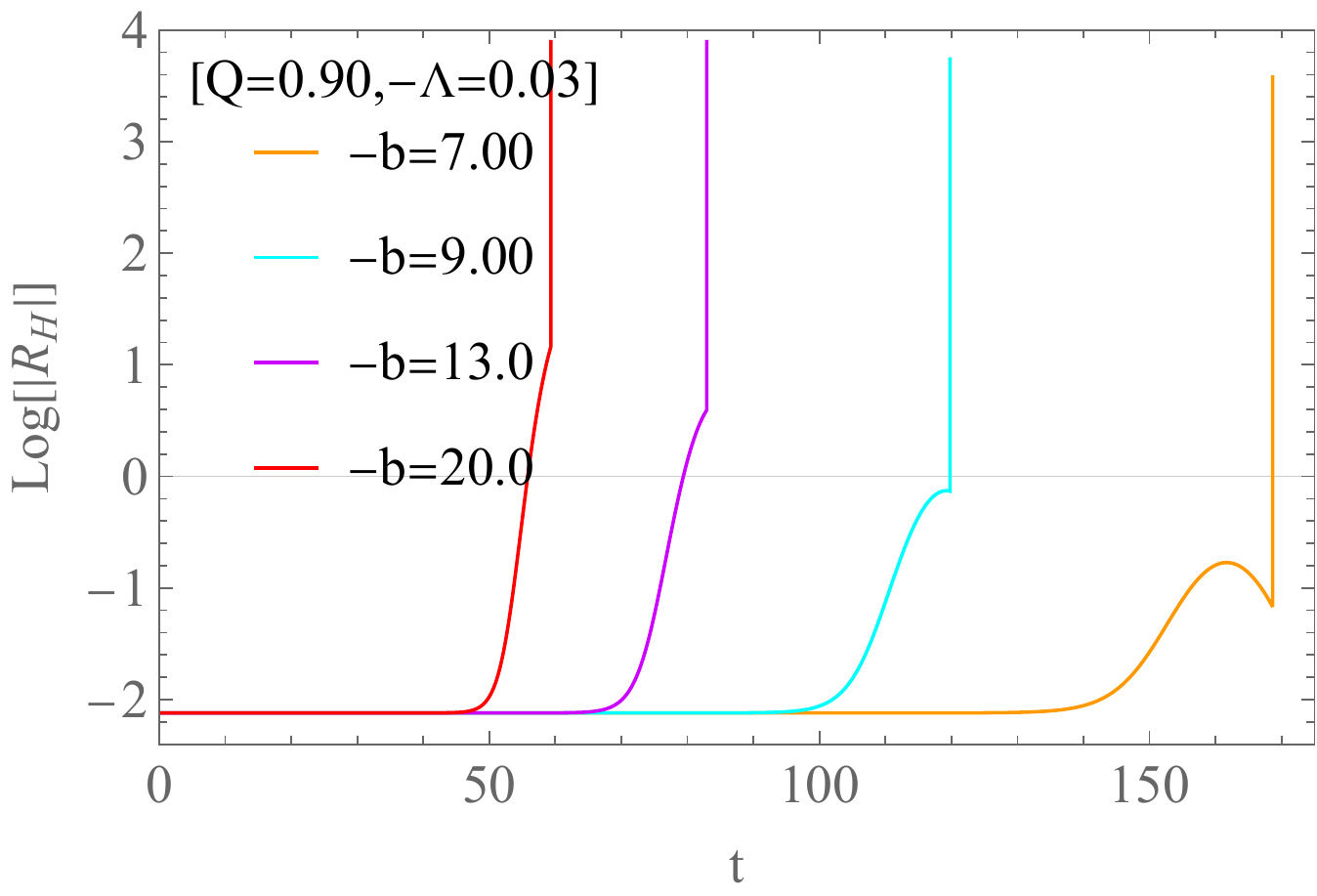}
    \end{minipage}
%    \begin{minipage}{0.3\linewidth}
% %       \centering
%       \includegraphics[width=4.8cm,height=4.5cm]{qdkjr}
%     \end{minipage}
    {\footnotesize{}\caption{{\footnotesize{}\label{fig:7}Left: the evolution of scalar curvature $R$ when $Q=0.9,-\Lambda=0.03$ and $-b=20$. The time step between adjacent curves is $\Delta{t}=1.8546$. The upper most curve corresponding to $t=59.3390$, after which our code crashes soon. The dashed parts represent  the results in the interior of the apparent horizon. Right:  the evolution of scalar curvature on the apparent horizon $R_H$ for various $b$ when $Q=0.9,-\Lambda=0.03$ before our code crashes.  %the  evolution of $\zeta$ on the apparent horizon.
    }}
    }{\footnotesize\par}
  \end{figure}
 The above subsection shows that the negative energy becomes more significant for stronger coupling parameter $-b$. Here we show that  $-b$ can not be too large, otherwise a naked singularity will appears inevitably. The left panel of Fig.~\ref{fig:7}  shows the evolution of the Ricci scalar for $Q=0.9,-\Lambda=0.03$ and $-b=20$. The Ricci scalar explodes  in the interior of the apparent horizon. Although our code  crashes at late times, we suggest that the curvature singularity moves outwards rapidly and finally passes through the apparent horizon such that a naked singularity forms. From another viewpoint, we show the evolution of the scalar curvature on the apparent horizon $R_H$ in the right panel. The $R_H$ also explodes with time. For larger $-b$, the $R_H$ increases faster and our code crashed earlier. We conclude that for large $-b$, the evolution endpoint of a linearly unstable RN-AdS black hole is a spacetime with naked singularity such that the  weak cosmic censorship is violated \cite{Wald:1997wa}. The cosmic censorship has been tested in EMS theory \cite{Corelli:2021ikv}, in which they found that naked singularities do not form for certain coupling functions. We will show later that for hyperbolic and power couplings,  naked singularities also do not form. In eSTGB theory, the  the cosmic censorship has been tested  very recently \cite{Corelli:2022pio,Corelli:2022phw}. They simulated the mass loss due to evaporation at the classical level using an auxiliary phantom field and suggested that either the weak cosmic censorship is violated or   horizonless remnants are produced. Here we find that without introducing phantom field, the cosmic censorship can also be violated. 

 \subsubsection{Irreducible mass of fractional coupling}
 
\begin{figure}[htbp]
  \begin{minipage}{0.3\linewidth}
    \centering
    \includegraphics[width=4.8cm,height=9cm]{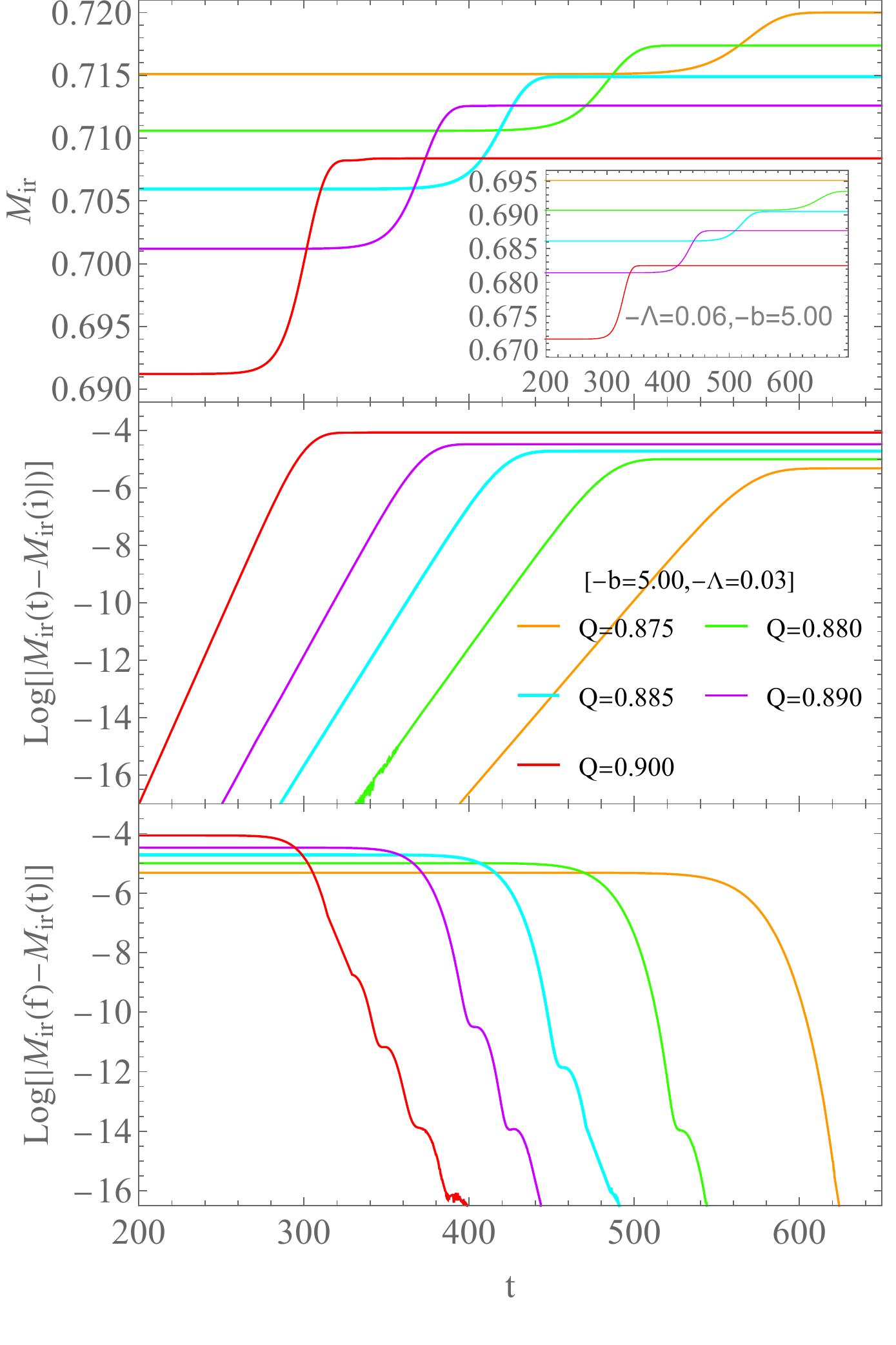}
  \end{minipage}
\begin{minipage}{0.3\linewidth}
   \centering
   \includegraphics[width=4.8cm,height=9cm]{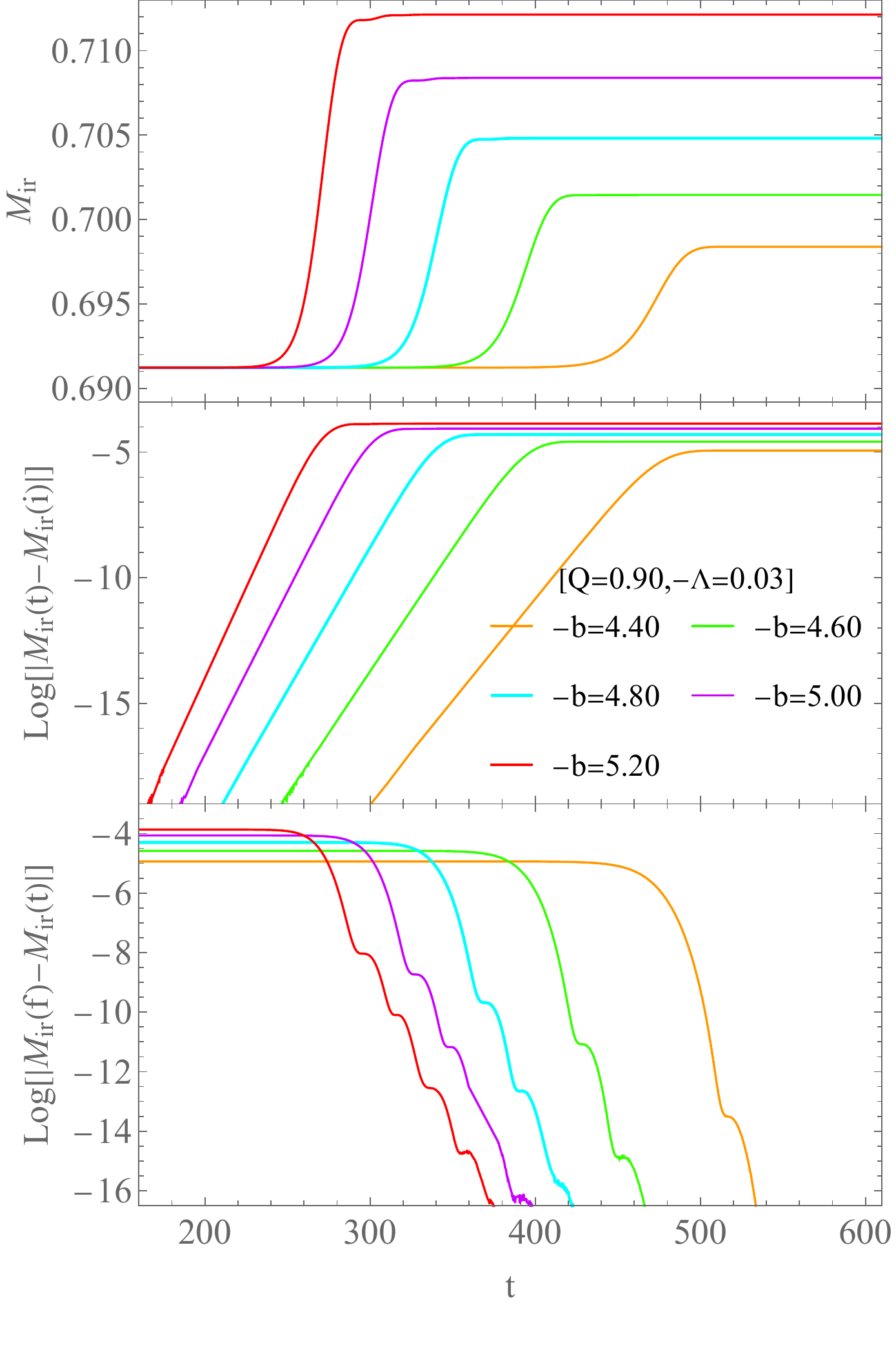}
  \end{minipage}
\begin{minipage}{0.3\linewidth}
   \centering
   \includegraphics[width=4.8cm,height=9cm]{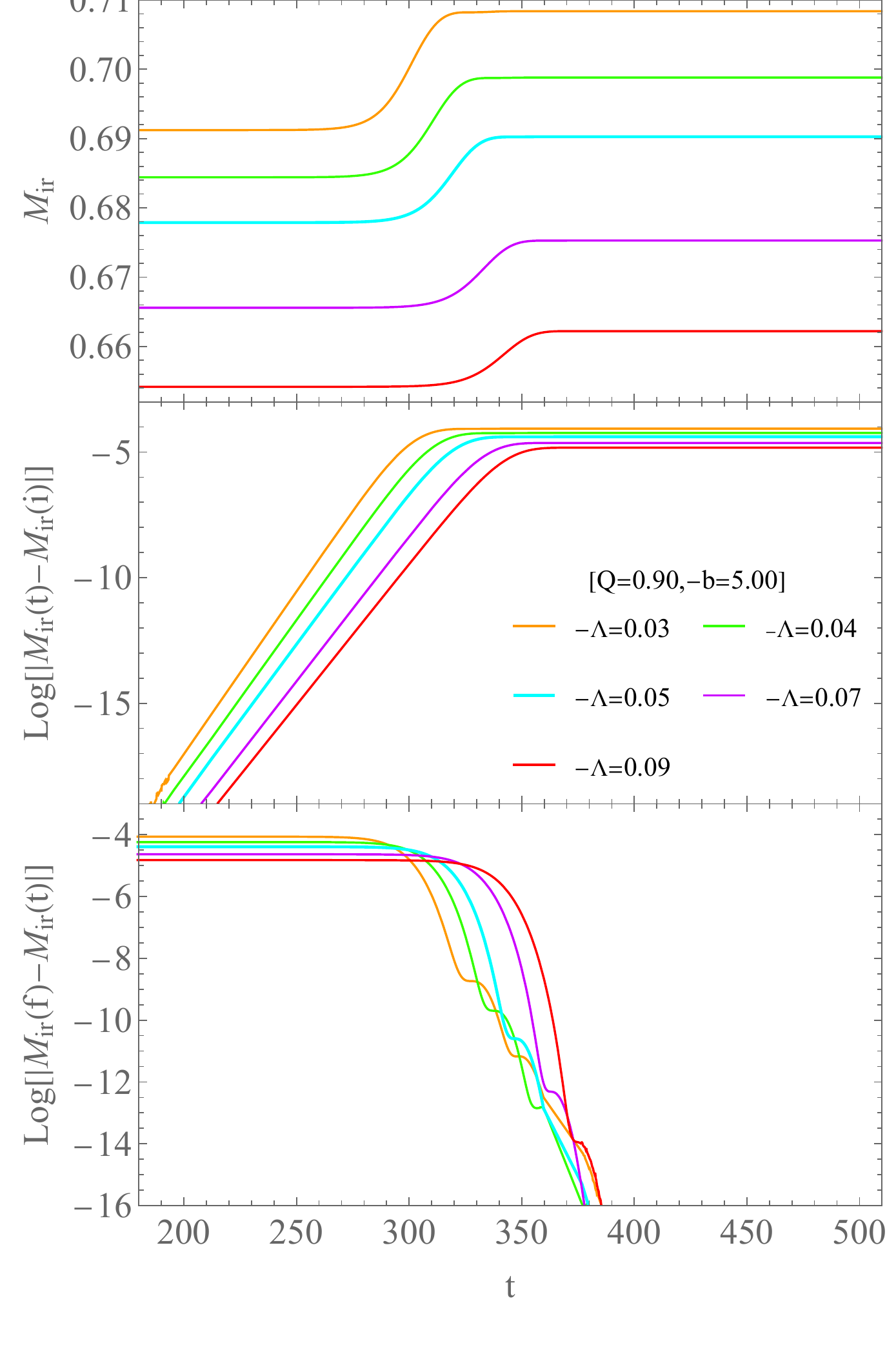}
\end{minipage}
  {\footnotesize{}\caption{{\footnotesize{}\label{fig:3}The evolution of irreducible mass $M_{ir}$ for various coupling constant $b$, charge $Q$ and cosmological constants $\Lambda$. %Since entropy is proportional to the horizon area ($M_{ir}$ is the horizon area from eq.~(\ref{eq:t})), we find that fractional coupling is thermodynamically preferred in the EMS gravity model.
  }}
  }{\footnotesize\par}
\end{figure}

Fig.~\ref{fig:3} displays the evolution of the BH irreducible mass $M_{ir}$ for various $Q,-b,-\Lambda$. The irreducible mass equals the BH apparent horizon area radius. In the upper row one can find that the irreducible mass never decreases during the evolution, although the weak energy condition is violated, as discussed in the above subsection. This is permissible since the the weak energy condition is a sufficient but not necessary condition for the black hole area increase law \cite{HawkingEllis,Nielsen:2008cr}. The nonlinear evolution exhibits no other obvious pathologies apart from the negative energy density. The scalarized solutions  are both thermodynamically and dynamically preferred.

%ensures that the surface area of a  BH   never decreases. Thus the scalarized BH solution bifurcating from the scalar-free BH solution is dynamically preferred. On the other hand, the BH thermodynamics claims that the BH entropy is proportional to the horizon area, and the second law of thermodynamics states that the entropy of a system never decreases, so 
 
The irreducible mass increases with $Q$ and $-b$. This can be understood from the coupling term between the Maxwell field and the scalar field in the action. For larger $Q$ or $-b$, the coupling  is stronger. More energy will be transferred from the Maxwell field to the scalar field. The BH can swallow more scalar field and its area grows. %From the upper plane of Fig.~\ref{fig:2}, the relationship between $\phi_{H}$ and $Q$ or $b$ implies that the strength of the repulsive effect is approximately linear with $Q$ or $b$ (since $\phi_{H}$ does not vary too significantly with $Q$ and $b$).
The cosmological constant $\Lambda$, however, puts more stringent condition  for the spontaneous scalarisation. Comparing the evolution of $M_{ir}$ at different $\Lambda$ in the upper-left inset of Fig.~\ref{fig:3}, within certain parameter ranges, the original  scalar-free BH is stabilized due to the increase of $-\Lambda$. In fact, in asymptotic AdS spacetime, the tachyonic instability occurs only when its effective mass-squared is less than the Breitenlohner-Freedman bound $\mu^{2}_{BF}=\frac{3\Lambda}{4}$ \cite{BFbound,Zhang:20211,Guo:2021}. For large enough $-\Lambda$, the tachyonic instability can be quenched. 
 
%The large the $-\Lambda$, the deeper and narrower the potential well. Less energy will be transferred from the Maxwell field to the scalar field. Therefore, we can find that the gap between the irreducible mass of the scalarized BH solution and the original scalar-free RN-Ads BH solution decreases with $\Lambda$.

 Another interesting feature is that the evolution of irreducible mass $M_{ir}$ can be roughly divided into two stages. The center and lower rows of Fig.~\ref{fig:3} illustrate that both the early stage and the late stages  follow   exponential evolution: %. The evolution of  $M_{ir}$ can be approximated by 
 \begin{equation}
  M_{ir}(t)\approx\begin{cases}
    M_{ir}(i)+\exp(\gamma_{i}{t}+\gamma_{1}), & \text{early times,}\\
    M_{ir}(f)-\exp(-\gamma_{f}{t}+\gamma_{2}), & \text{late times.}
  \end{cases} 
 \end{equation}
 Here $\gamma_{i}$ and $\gamma_{f}$ are the exponential growth rate and saturation rate of $M_{ir}$, respectively.  $M_{ir}(i)$ and $M_{ir}(f)$ are the initial and final irreducible mass of the BH, respectively. $\gamma_{1,2}$ are some terms less important. Note that $M_{ir}(i)$ of the initial RN-AdS BH depends on $Q$ and $\Lambda$.  %, and $M_{ir}(f)$ of scalarized BH is related to all the fundamental parameters. 
 From the middle row of Fig.~\ref{fig:3}, the relationship between  $\gamma_{i}$   and  $Q,b,\Lambda$ is analogous to those of the $\phi_{H}$ at the horizon. However, the saturation stage is stepped rather than damped oscillation, as shown in the lower row of Fig.~\ref{fig:3}.

\subsubsection{$\phi_{3}$ of fractional coupling}
\begin{figure}[htbp]
  \begin{minipage}{0.3\linewidth}
    \centering
    \includegraphics[width=4.8cm,height=8cm]{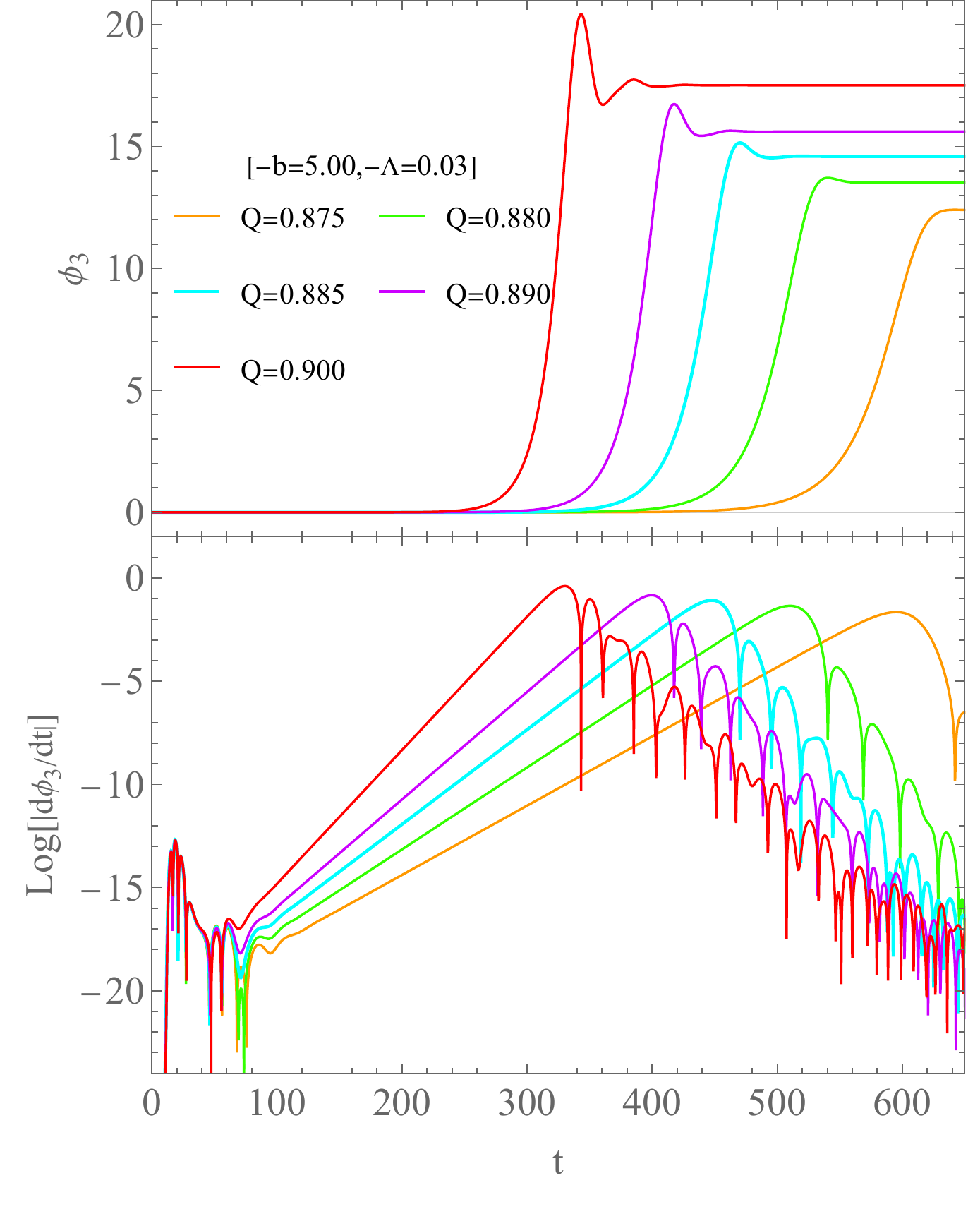}
  \end{minipage}
  \begin{minipage}{0.3\linewidth}
    \centering
    \includegraphics[width=4.8cm,height=8cm]{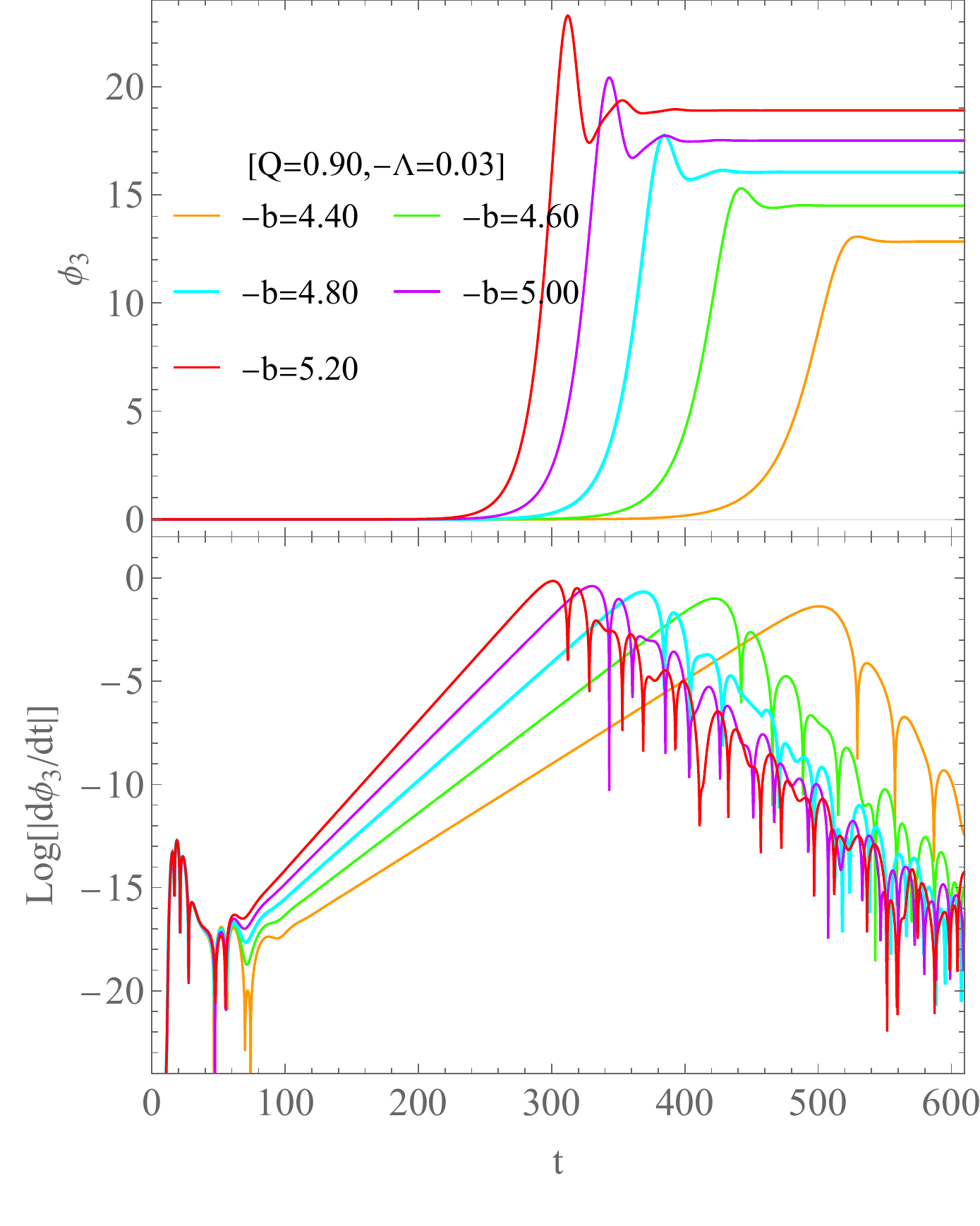}
  \end{minipage}
  \begin{minipage}{0.3\linewidth}
    \centering
    \includegraphics[width=4.8cm,height=8cm]{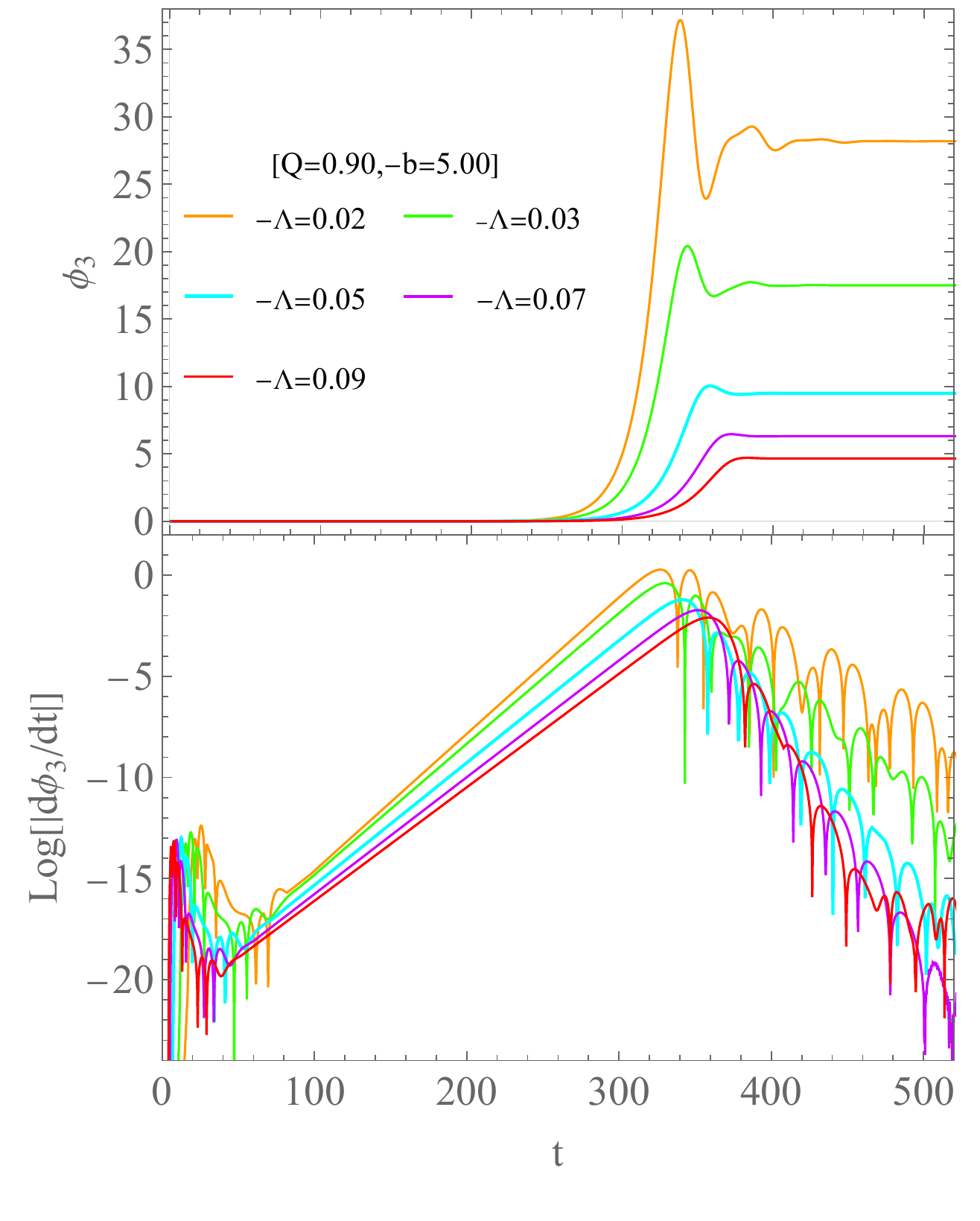}
  \end{minipage}
  {\footnotesize{}\caption{{\footnotesize{}\label{fig:6}The evolution of $\phi_{3}$. It resembles the evolution of   $\phi_{H}$ , although the sign is reversed (upper row). The lower row shows the evolution of $\log|\frac{d\phi_{3}}{dt}|$.  %From the lower row, $Q$, $b$ and $\Lambda$ control the time for the system to settle down by affecting the imaginary frequency $\mu_{\phi}$.
  }}
  }{\footnotesize\par}
\end{figure}

Now we investigative the evolution of coefficient $\phi_{3}$ of the scalar field at spacial infinity.   Fig.~\ref{fig:6} shows that the evolution of $\phi_{3}$ resembles the evolution of $\phi_{H}$,  which can also be divided roughly into two stages. At early stage, it increases exponentially. At late time, it converges to the equilibrium value $\phi_{3}(f)$  with damped oscillation which resembles the quasinormal mode. Its evolution can be approximated by 
 \begin{equation} 
 \phi_{3}\approx\begin{cases}
   \exp(\eta_i{t}+\eta_1), & \text{early times,}\\
   \phi_{3}(f)-\exp(-\eta_f{t}+\eta_2), & \text{late times.}
  \end{cases} 
 \end{equation} 
 Here $\eta_i$ is the grows rate of $\phi_3$ at early times, and $\eta_f$   the imaginary part of the dominant mode frequency of $\phi_3$ at late times. $\eta_{1,2}$ are some terms less important. The lower row of   Fig.~\ref{fig:6} shows that $\eta_i$ is positively related to $Q,-b$ and negatively related to $\Lambda$. Meanwhile $\eta_f$ has contrary relations to $Q,-b,\Lambda$. %When $Q$ and $-b$ increase, the oscillation is more intense. Coversely, a sufficiently large $-\Lambda$ will suppress the oscillation of $Log[|\frac{d{\phi_{3}}}{dt}|]$, which leads to a shortened convergence time of the final state. %However, one thing in common is that $Log[|\frac{d{\phi_{3}}}{dt}|]$ is difficult to converge to a constant in these fundamental parameters.

 %Compare carefully, the final state imaginary frequency $\mu_{\phi}$ of $Log[|\frac{d{\phi_{3}}}{dt}|]$ has the same regular change as $Log[|\phi_{H}(f)-\phi_{H}(t)|]$, $i.e.$ the absolute value of $\mu_{\phi}$ is inversely proportional to $Q$ and $-b$. In fact, this is the inevitable result of the competition between the gravitational and repulsive effects of BHs. As long as one of them plays a dominant role, the time for the system to settle down is relatively short. If the former and the latter compete together, the system will difficulty converging to equilibrium. All of these outcomes depend on $Q$ and $b$. 

There are universal and robust relationships between $\phi_{H}$, $M_{ir}$ and $\phi_{3}$ during the evolution: 
 \begin{equation}
     \gamma_{i}=2\nu_i=2\eta_i, \gamma_{f}=2\nu_f=2\eta_f.\label{eq:mm}
 \end{equation}
 This relationship can be understood  for an intermediate solution  which can be approximated by a static solution.  For a static solution, the variables $S$, $\alpha$ and $P$ are zero on the horizon. So combining  (\ref{eq:h},\ref{eq:i},\ref{eq:j},\ref{eq:l}) and (\ref{eq:n}) one can find $\partial_{t}S$, $\partial_{t}\zeta\propto{\delta\phi^{2}}$ for the intermediate solution. Since $S(r_{H},t)=0$ and  (\ref{eq:t}) states that $M_{ir}=\zeta(r_H,t)$, we can deduce that $\Dot{M}_{ir}=\Dot{\zeta}(r_{H},t)=-\frac{\partial_t S}{\partial_r S}\partial_{r}\zeta+\partial_{t}\zeta|_{r_H}\propto{\Dot{\phi}_{H}^{2}}$. Since at early and late times, the evolution can be approximated by the perturbations for the initial  and   final BHs, respectively, this leads to the relations (\ref{eq:mm}). These relations have been found in other cases \cite{Zhang:2022,Zhang:20211,zhang:20221,zhang:20222}. % namely $\delta{M_{ir}}\propto{\delta{\phi_{H}}^{2}}$. %From energy perspective, Eq.~(\ref{eq:mm}) implies that the spontaneous scalarization of the BH leaves only the scalar field of lower order of magnitude outside the horizon.

\subsection{Results for Power-Law and Hyperbolic Coupling}
%\subsubsection{Solutions of Hyperbolic coupling $f_{H}(\phi)$ and power-law coupling $f_{P}(\phi)$}

\begin{figure}[htbp]
  \begin{minipage}{0.3\linewidth}
    \centering
    \includegraphics[width=4.8cm,height=8cm]{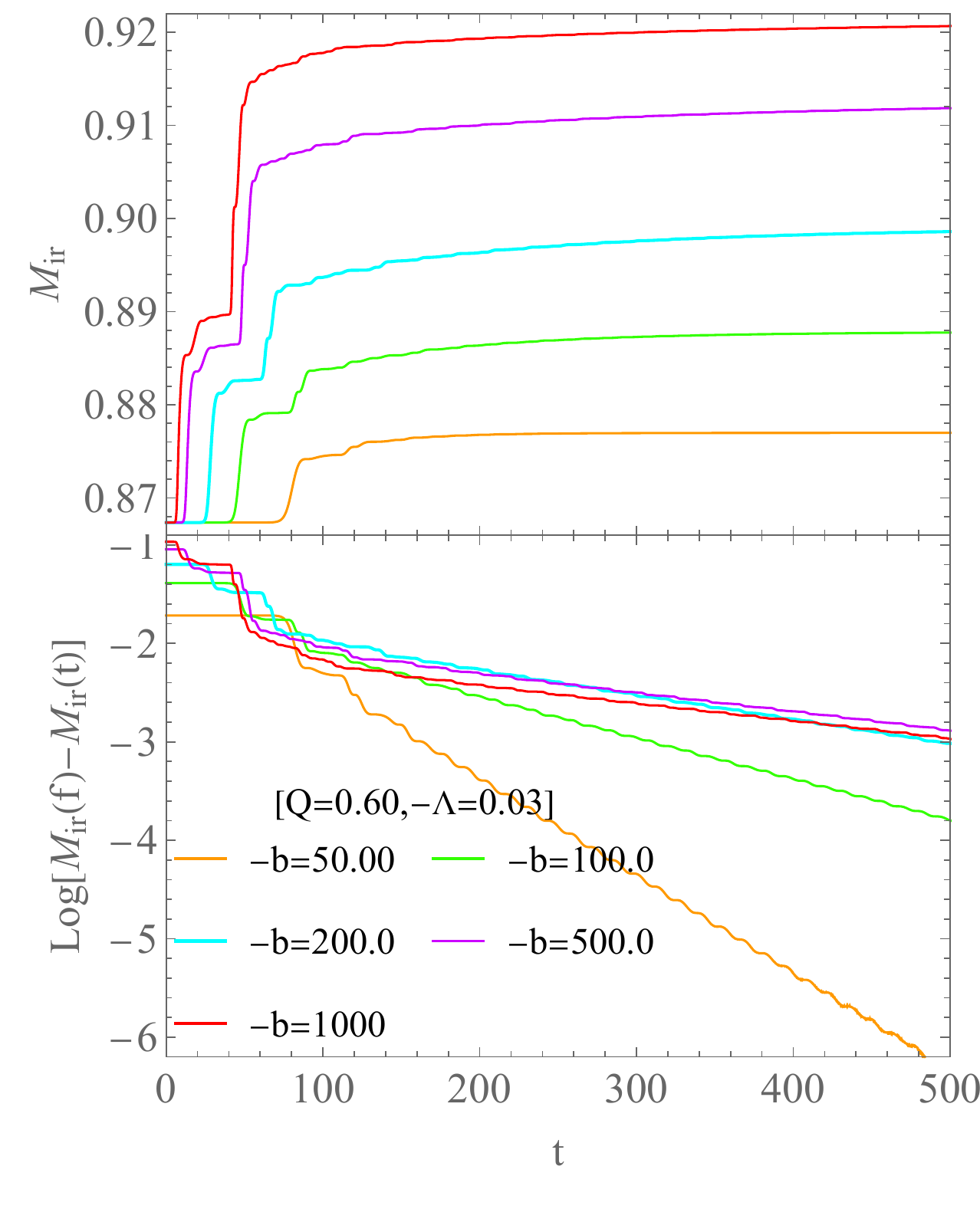}
  \end{minipage}
  \begin{minipage}{0.3\linewidth}
    \centering
    \includegraphics[width=4.8cm,height=8cm]{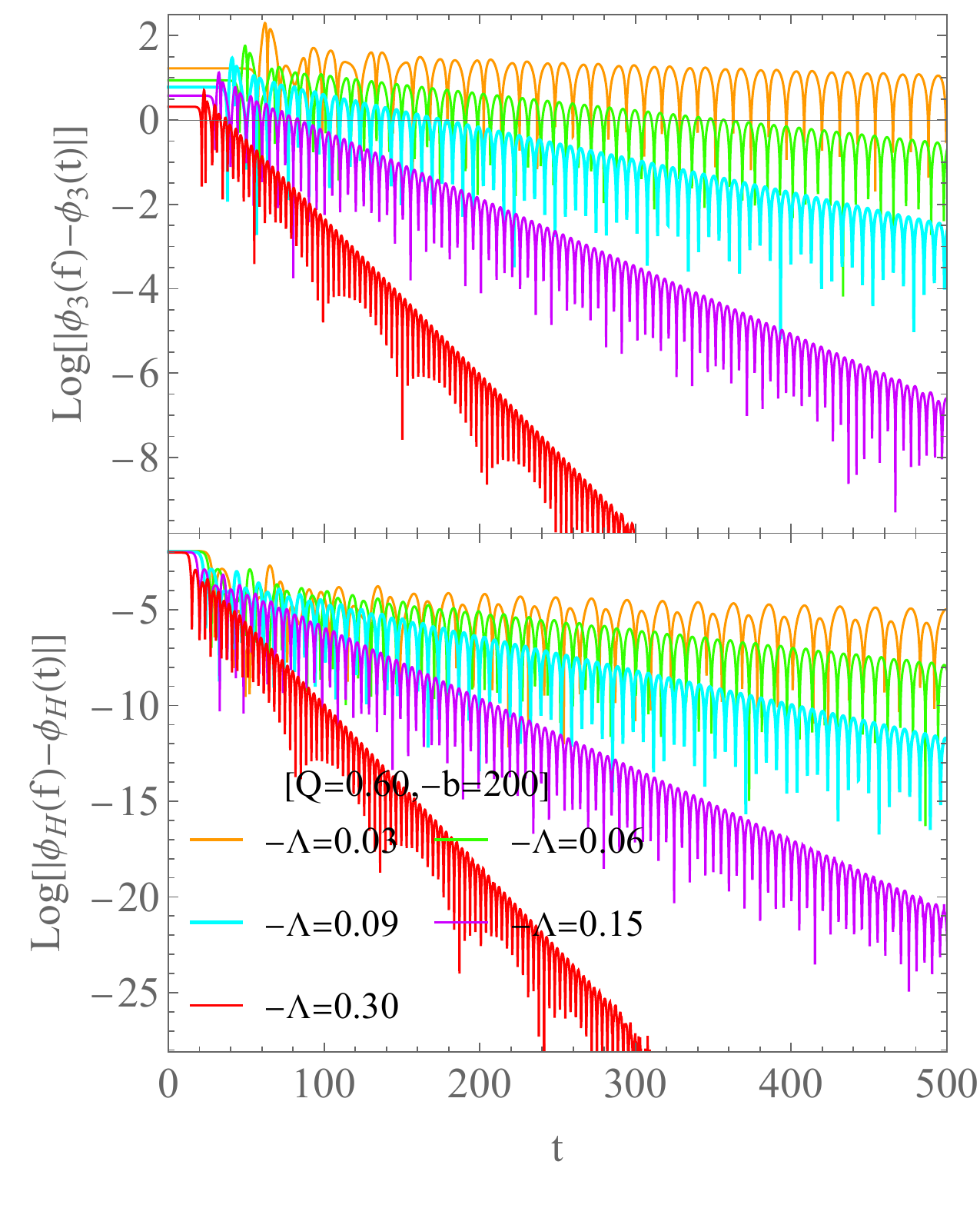}
  \end{minipage}
\begin{minipage}{0.3\linewidth}
  \centering
  \includegraphics[width=4.8cm,height=8cm]{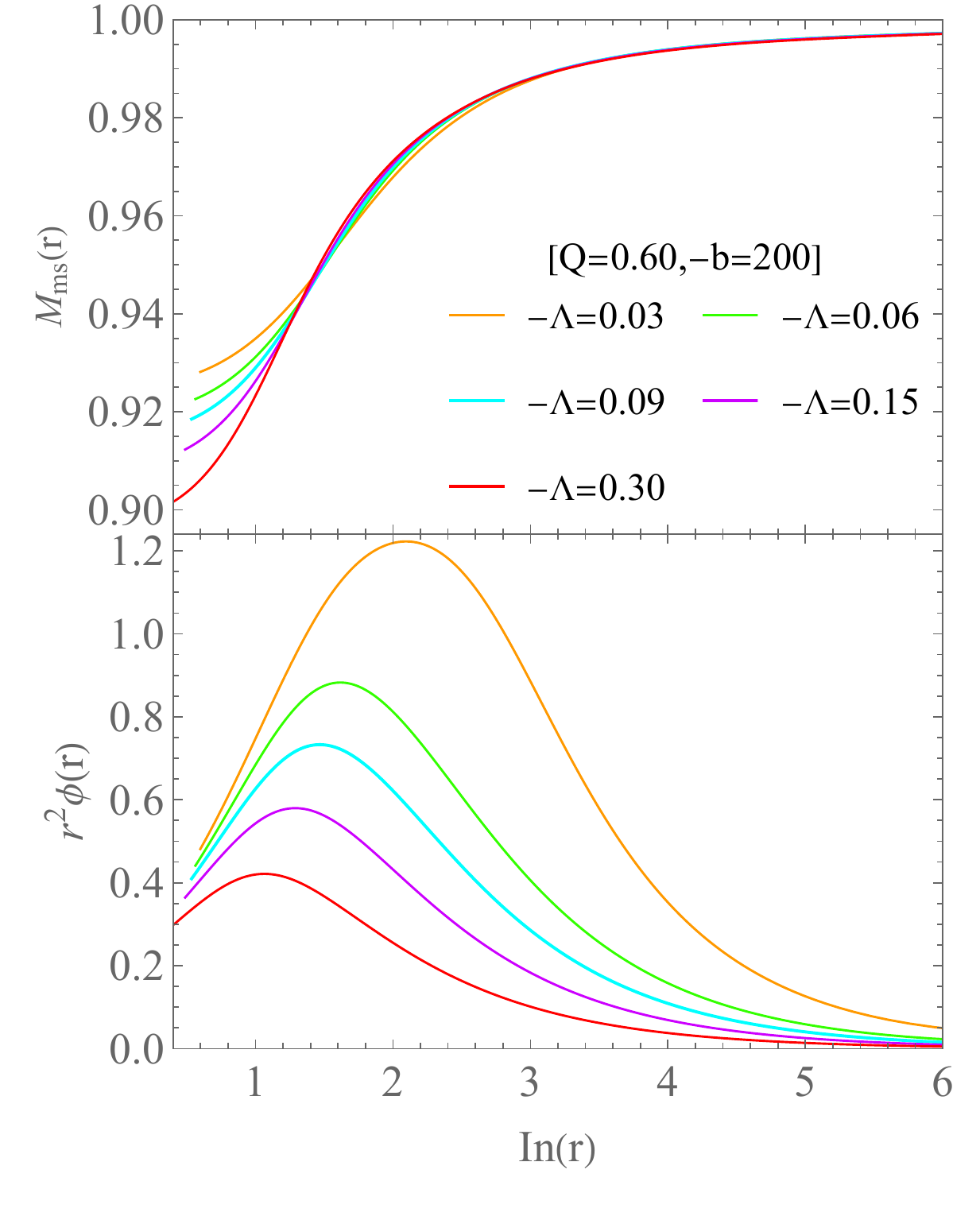}
\end{minipage}
  {\footnotesize{}\caption{{\footnotesize{}\label{fig:8}Left column: the evolution of the irreducible mass $M_{ir}$ for models with hyperbolic coupling function. Middle column: the evolution of $\phi_3$ and $\phi_H$.  
  Right  column: the final profiles of the Misner-Sharp mass  and $r^2\phi$ for different $\Lambda$.  %Right column: What's more, the intermediate evolution stage of $M_{ir}$ has many various steps compared to fractional coupling, but still obeys exponential growth. Interestingly, further studies revealed that the dynamical features of hyperbolic coupling, power-law coupling and exponential coupling are all consistent in the AdS-EMS gravity model. In fact, for $-b\phi^{2}<<1$}, they share the same first-order Taylor expansion, resulting in no divergence in the perturbation.}
  }}}  {\footnotesize\par}
\end{figure}

In this subsection, we consider the dynamics of the spontaneous scalarization with coupling functions $f_H=\cosh{\sqrt{-2b}\phi}$ and $f_P=1-b\phi^2$.  %The dynamic feature of hyperbolic coupling $f_{H}(\phi)$ in the AdS-EMS gravity model are more complex compared to fractional coupling $f_{F}(\phi)$. For example, the irreducible mass $M_{ir}$ has various step-like jump evolutions in ntermediate stage, 
The results are shown in Fig.~\ref{fig:8}. It can be seen that the evolution  of $M_{ir}$ and $\phi_H,\phi_3$ still obeys  the exponential growth at early stage and saturates to the equilibrium value at late times. But now  the Misner-Sharp mass  monotonically increases to the ADM mass at spacial infinity. In other words, there is no negative energy distribution outside the BH horizon. %Again, the oscillatory behavior of $Log[|\phi_{3}(f)-\phi_{3}(t)|]$ and $Log[|\phi_{H}(f)-\phi_{H}(t)|]$ are far more intense than discussed above.
We find that the dynamical features of the spontaneous scalarization for power-law  and hyperbolic coupling $f_E=\exp(-b\phi^2)$ in the AdS-EMS gravity model are  similar to the case with exponential coupling which has been studied in \cite{Zhang:20211}, and we will not repeat the text.  
 
\section{Conclusion}

We have focused attention on the dynamical spontaneous scalarization   in the asymptotically AdS spacetime in EMS models. We have discussed three different forms of coupling function, which share the same form of the leading-order expansion $1-b\phi^{2}$ in the limit of small scalar field.  
% In this model, a variety of scalarized solutions are possible, depending only on the coupling function that governs the non-minimal coupling between the source term and the scalar field. Subject to the constraints of choice of coupling function, we introduce the coupling in \cite{Fernandes:2019} in this context.
 Under certain conditions, 
 %studies have shown that RN-AdS BH are stable against scalar perturbations. For sufficiently large fundamental parameters, however, 
we have found that the bald RN-AdS BH can be transformed into scalarized BH, which is preferred in thermodynamics. 
% One of our motivations for this work is to figure out what effect the fundamental parameters have on the physical properties scalarized BHs.
We  have explored the effects of the BH charge $Q$, coupling strength parameter $b$ and cosmological constant $\Lambda$ on the dynamical process in the scalarization. When the system reaches equilibrium, the extreme value of $\phi$ always locates at the horizon (denoted as $\phi_{H}$). % and is positively related to $Q$ or $-b$.
Starting from the initial bald RN-AdS BH,  we find that $\phi_{H}$ grows exponentially at the early stage of the dynamical evolution in the scalarization. % and the growth coefficient $\gamma$ increases with $Q$ and $-b$. 
At the late stage in the process of scalarization, $\phi_{H}$ converges to an equilibrium value through  damped oscillation.
We find that the scalarization is enhanced by larger values of $Q,-b$ but suppressed with the increase of $\Lambda$. 
%The reason is that the spontaneous scalarization is triggered by the tachyonic instability, which is enhanced by  $Q,-b$ but suppressed by $\Lambda$ in AdS spacetime. 
We hae also investigated the evolution of  $\phi_{3}$, %One motivation is to explore the connection between the time for the system to settle down and the fundamental parameters. 
and find   $\phi_{3}$ evolves similarly to $\phi_{H}$. % even though their initial behavior is different and sign diverge. Further research interprets that $Q$, $b$ and $\Lambda$ change the evolution time of the system by affecting the imaginary frequency $\mu_{\phi}$ of the damped oscillation. In fact, since $Q$ and $b$ control the strength of repulsive effect, the increase in $Q$ and $-b$ grants the repulsive effect the power to fight against gravity, thereby causing the change in $\mu_{\phi}$. As for $\Lambda$, it acts like a potential well so that the absolute value of $\mu_{\phi}$ keeps increasing with $-\Lambda$.

 The irreducible mass $M_{ir}$   never decreases during the dynamical spontaneous scalarization of the BH. Since  $M_{ir}$ is the horizon area radius and the BH entropy is proportional to the horizon area, this feature is a signal that  the second law of thermodynamics is obeyed, although the weak energy condition is violated in models with fractional coupling function.  $M_{ir}$  grows exponentially at early times and saturates also exponentially to the final value at late times. The corresponding growth coefficient $\gamma_{i}$ and saturation coefficient $\gamma_{f}$    increase with $Q$ and $-b$. %but they have different physical effects, that is, 
 The increase of $\gamma_{i,f}$ can shorten the growth and saturation time of $M_{ir}$. %, while $\gamma_{f}$ will prolong the saturation time of $M_{ir}$.
 On the other hand, the cosmological constant plays a contrary role that prolongs the time for dynamical scalarization.
 %the stability of the initial RN-AdS BH is stronger for larger $-\Lambda$, which resulted in the growth and saturation coefficients having opposite relations compared to $Q$ and $-b$. %But the irreducible mass is still thermodynamically preferred under the constraint of $\Lambda$.

For EMS model with fractional coupling,  there is negative energy distribution  near the BH horizon such that the weak energy condition is violated. 
%If the values of $Q$, $b$ and $\Lambda$ are appropriate so that $1+b\phi^{2}<0$, the Minsner-Sharp mass $M_{ms}$ will be divided into two continuous monotone regions, $i.e.$ decreasing monotonically first and then asymptotically converging to the ADM mass $M=1$ at radial infinity. The existence of multiple monotonic regions in $M_{ms}$ is due to the appearance of negative energy bands $\Delta{z}$ in the vicinty of horizon.
The negative energy region is stretched with the increase of $Q$ and $-b$ and narrowed with $\Lambda$.  
% and the negative energy density $\rho$ in it is also growing, resulting in the expansion of the decreasing region of $M_{ms}$ and forcing $M_{ms}(min)$ away from the horizon. But the cosmological constant $\Lambda$ is averse to negative energy, and its reduction on the one hand inhibits the energy flux flowing into the BH, and on the other hand pushes $\rho$, which violates the energy condition, into the horizon. Therefore, the negative energy band $\Delta{z}$ and the negative energy density $\rho$ contract as $-\Lambda$ increases.
However, $Q$ and $-b$  cannot be too large. Once these   parameters reach maximum thresholds, a naked singularity will appears and the weak cosmic censorship is violated. 
%the metric function  $\zeta$ becomes zero and our numerical   codes crash.  % Therefore, we cannot investigate the dynamical evolution of BHs after the roles of gravitational and repulsive effects are reversed. %For the relationship between $\mu_{\phi}$ and $\Lambda$, research shows that $\mu_{\phi}$ will approach a non-zero constant as $\Lambda\to0$. In AdS spacetime, $\Lambda$ approaches zero, which means that the potential well asymptotically vanishes. As a consequence, the time required for scalar field to traverses the entire spacetime will be extended. For the growth coefficient $\eta$ in the intermediate stage of $Log[|\frac{d{\phi_{3}}}{dt}|]$, the relationship between it and free parameters $Q$, $b$ and $\Lambda$ is also similar to $\gamma$ and $\gamma_{i}$.
Compared with fractional coupling, the cases with hyperbolic coupling $f_{H}(\phi)$ and power-law coupling $f_{P}(\phi)$ in AdS spacetime do not violate   energy condition. Note that $f_{H}(\phi)$, $f_{P}(\phi)$ and exponential coupling $f_{E}(\phi)$ share the same leading order expansion in perturbations. %This leads to their consistent dynamics.
Therefore, the dynamic evolution features of $f_{H}(\phi)$ and $f_{P}(\phi)$ can be similar to  those with the exponential coupling $f_{E}(\phi)$ \cite{Zhang:20211}. 

\section*{Acknowledgments}
This work is supported by the Natural Science Foundation of China under Grant No. 11805083, 11905083, 12005077, 12075202 and Guangdong Basic and Applied Basic Research Foundation (2021A1515012374).

\end{document}